\documentclass[11pt]{JHEP}
\usepackage{amsmath}
\usepackage{amssymb}
\usepackage{amsmath,amsthm,epsfig,euscript,array,cancel}
\font\mybb=msbm10 at 10pt
\def\bb#1{\hbox{\mybb#1}}

\newcommand{\be}{\begin{equation}}
\newcommand{\ee}{\end{equation}}

\def\bea{\begin{eqnarray}}
\def\eea{\end{eqnarray}}
\preprint{
V2. November 14, 2019
}
\renewcommand{\theequation}{\arabic{section}.\arabic{equation}}
\title{On polarized scattering equations for superamplitudes
of 11D supergravity and ambitwistor superstring}

\author{ Igor Bandos
\\ {\small\it Department of
Theoretical Physics, University of the Basque Country UPV/EHU, \\ P.O. Box 644, 48080 Bilbao, Spain} \\
{\small\it and IKERBASQUE, Basque Foundation for Science, 48011, Bilbao, Spain}
}

\date{Jan 22- August 20 - October 21, 2019,  Printed \today }

\abstract{
We revisited the formalism of 11D polarized scattering equation by Geyer and Mason from the perspective of spinor frame approach and spinor moving frame formulation of the 11D ambitwistor superstring action.
In particular, we rigorously obtain the equation for the spinor  function on Riemann sphere from the supertwistor form of the ambitwistor superstring action, write its general solution and use it to derive the polarized scattering equation. We show that the expression  used by Geyer and Mason to motivate their ansatz for the solution of polarized scattering equation can be obtained from our solution after a suitable gauge fixing. To this end we use the hidden gauge symmetries of the 11D ambitwistor superstring, including $SO(16)$, and the description of ambitwistor superstring as a dynamical system in an 11D superspace enlarged by bosonic directions parametrized by 517 tensorial central charge coordinates $Z^{\underline{\mu} \underline{\nu}}$ and $Z^{\underline{\mu}\underline{\nu}\underline{\rho}\underline{\sigma}\underline{\kappa}}$.
\\
We have also found the fermionic superpartner of the polarized scattering equation. This happens to be a differential equation in fermionic  variables imposed on the superamplitude, rather then just a condition on the scattering  data as the bosonic polarized scattering equation is.
\\ D=10 case is also discussed stressing the  similarities and differences with 11D systems. The useful formulation of 10D ambitwistor superstring  considers it as a dynamical system in  superspace enlarged with 126 tensorial central charge coordinates $Z^{\mu\nu\rho\sigma\kappa}$.}

\keywords{supersymmetry, supergravity, amplitudes, twistor approach, higher dimensions, spinor moving frame}

\begin{document}

\section{Introduction}

Recent years  an impressive progress in calculation of scattering amplitudes of maximally supersymmetric theories was reached
 \cite{Bern:2011qn,Drummond:2008vq,Drummond:2009fd,Eden:2011ku,Kallosh:2012yy,Elvang:2015rqa,ArkaniHamed:2017,Caron-Huot:2019vjl}. It is related mainly with the use of on-shell methods, in particular of spinor helicity formalism (closely related to twistor approach \cite{Penrose:1967wn,Penrose:1972ia,Atiyah:2017erd}) and its superfield generalization
\cite{Britto:2005fq,Bianchi:2008pu,ArkaniHamed:2008gz,Brandhuber:2008pf,Mason:2009qx,Heslop:2016plj,Herrmann:2016qea} which is especially simple and  efficient in the case of 4 spacetime dimensions.

The development of this twisor-like formalism for the case of higher dimensional theories and its applications were discussed in
\cite{Cheung:2009dc,CaronHuot:2010rj,Boels:2012ie,Boels:2012zr,Wang:2015jna,Wang:2015aua,Bandos:2016tsm,Bandos:2017zap,Bandos:2017eof,Geyer:2019ayz}.
In particular, in \cite{Bandos:2016tsm,Bandos:2017eof} the observation that 10D spinor helicity formalism of \cite{CaronHuot:2010rj} can be understood as spinor moving frame approach to supersymmetric particles extended to the description of amplitudes\footnote{See \cite{Uvarov:2015rxa} for similar observation in 5d context. The above references deal mainly with the case of flat spacetime/superspace.  Twistor methods for AdS$_5$ and AdS$_5\times {\bb S}^5$ were addressed e.g. in \cite{Adamo:2016rtr} and \cite{Uvarov:2019vmd}. Spinor helicity formalism for AdS scattering amplitudes \cite{Heemskerk:2009pn} (see also \cite{Albayrak:2019asr} and references therein) was the subject of recent \cite{Nagaraj:2018nxq}.} allowed us to develop the spinor helicity formalism for 11D supergravity and a new constrained superfield formalism for 10D SYM and 11D supergravity amplitudes, to find the Ward identities for these amplitudes and to
discuss a candidate for generalization of the BCFW  recurrent relations \cite{Britto:2005fq} for
the constrained tree superamplitudes. In    \cite{Bandos:2017zap} an alternative analytic superfield formalism for superamplitudes was proposed.
It was also oriented on the use of BCFW--type  recurrent relations which are still to be found in this case.

More recently an apparently different  approach to 11D supergravity and 10D SYM amplitudes was proposed in \cite{Geyer:2019ayz}. It is based on the so-called {\it polarized scattering equation} which can be considered as a kind of square root of the
CHY  scattering equations \cite{Cachazo:2013hca,Cachazo:2014xea} (actually present already in \cite{Gross:1987kza,Gross:1987ar,Gross:1989ge}; see \cite{Cachazo:2019ngv} for recent development and more references).
The polarized scattering equation for 6D amplitudes was proposed in \cite{Geyer:2018xgb} while the 11D and 10D polarized scattering equations are among the beautiful findings of  \cite{Geyer:2019ayz}. Its  relation with ambitwistor string models \cite{Mason:2013sva,Adamo:2013tsa,Bandos:2014lja,Geyer:2014fka,Lipstein:2015vxa,Casali:2016atr,Casali:2017zkz,Adamo:2018ege,Carabine:2018kdg}, the 11D and 10D versions of which were considered for the first time in \cite{Bandos:2014lja},
was discussed and especially stressed in \cite{Geyer:2019ayz}.

In this paper we revisit the 11D polarized scattering equation formalism of \cite{Geyer:2019ayz} and its ambitwistor superstring origin using the spinor frame approach. We show how the understanding of the spinor frame nature of the 11D spinor helicity formalism allows to clarify the origin of basic equations imposed in
 \cite{Geyer:2019ayz} and the ambitwistor superstring derivation of these equations.
We show that the correct basis for this is provided by the 11D ambitwistor superstring of \cite{Bandos:2014lja}
rather then by its modification suggested in \cite{Geyer:2019ayz}. In the derivation of the basic equations the solution of which provides us with the expression for the meromorphic spinor function, which was employed to formulated the polarized scattering equation in its most suggestive form, we have used essentially the possibility to formulate the 11D ambitwistor superstring as   a system in an enlarged superspace with 528 bosonic coordinates
 \cite{Bandos:2014lja} as well as the $SO(16)$ gauge symmetry of the 11D ambitwistor superstring\footnote{The authors of  \cite{Geyer:2019ayz} proposed a modification of the twistor form of the ambitwistor superstring action of \cite{Bandos:2014lja}  by reducing this $SO(16)$ to $SO(13)$ gauge symmetry.
 }.

In the  ambitwistor superstring approach the above mentioned meromorphic spinor function on Riemann sphere, which satisfies the polarized scattering equation, appears accompanied by 16 component fermionic meromorphic function. The expression for this in terms of scattering data is supersymmetric invariant after the expression for meromorphic bosonic spinor function is taken into account and thus can be considered as a superpartner of this latter. This observation suggests the existence of a fermionic superpartner of the polarized scattering equations. We show that such a superpartner ({\it spolarized scattering equation}) does exist but is a differential equation satisfied by 11D  superamplitudes rather than a condition on scattering data (as the bosonic polarized scattering equation is itself).

We also consider 10D polarized scattering equation formalism and its ambitwistor superstring origin especially stressing the stages where the differences with 11D case occur.

We begin in sec. 2 by reviewing the spinor frame description of the 11D spinor helicity formalism  \cite{Bandos:2017zap}. In sec. 3, after reviewing the scattering equation \cite{Cachazo:2013hca} (sec. 3.1), we revisit the 11D polarized scattering equation of \cite{Geyer:2019ayz} with the use of  the spinor frame version of the spinor helicity formalism \cite{Bandos:2016tsm,Bandos:2017zap,Bandos:2017eof}. We show there how the polarized scattering equation appears as consistency condition of the constraints for the meromorphic spinor function and scattering equation for the meromorphic vector function involved in these constraints.  In sec. 4 we reconsider the supersymmetry generator and supersymmetric amplitude proposed in \cite{Geyer:2019ayz} from the perspective of spinor frame approach.
In Sec. 5 we turn to the ambitwistor superstring origin of the polarized scattering equation. We begin there by briefly reviewing the standard Green-Schwarz/Brink-Schwarz like formulation of the ambitwistor superstring and its reformulation in term of constrained supertwistor $(\mu^{\underline{\alpha}}_{\underline{q}}, \lambda_{\underline{\alpha}\underline{q}}, \eta_{\underline{q}})$.
We show that the fact that 11D ambitwistor superstring of \cite{Bandos:2014lja}  can be formulated as a dynamical system in an enlarged superspace can be used to relax the  second class constraints restricting $\mu^{\underline{\alpha}}_{\underline{q}}$. Then, applying the Lagrange multiplier method we can introduce  the first class constraints generating $SO(16)$ gauge symmetry into the supertwistor form of the ambitwistor superstring action and consider  the supertwistor component $\mu^{\underline{\alpha}}_{\underline{q}}$  as unconstrained variable. The variation of this first order action with respect to $\mu^{\underline{\alpha}}_{\underline{q}}(\sigma)$ becomes straightforward and
is used to obtain the dynamical equation for highly constrained bosonic fields, the  spinor functions $\lambda_{\underline{\alpha}\underline{q}}(\sigma)$. The solution of these equations provides us with the $SO(16)$ gauge covariant generalization of the meromorphic spinor functions used as an ansatz for the solution of the polarized scattering equation in \cite{Geyer:2019ayz}.

In sec. \ref{sec=fpScEq} we find the fermionic superpartner of the polarized scattering equation which is a differential equation imposed on superamplitudes.
Finally, in sec. 7 we describe briefly the D=10 spinor helicity formalism, polarized scattering equation and
ambitwistor superstring origin of this, especially stressing the points where the 10D case differs from 11D one.
We conclude in Sec. 8. Some useful equations of the spinor frame formalism can be found in the Appendices.

Our notation are that of \cite{Bandos:2016tsm,Bandos:2017eof} and \cite{Bandos:2017zap}, up to the use of underlined Greek symbols from the beginning/middle of the alphabet for the 11D Majorana spinors/vectors and underlined Latin symbols for the SO(9) vector ($\underline{I}, \underline{J},...$)  and spinor indices ($\underline{q}, \underline{p},...$). In some places $\underline{q}, \underline{p},...$ are also considered to be $SO(16)$ vector indices, which is related to the hidden SO(16) symmetry of 11D superparticle (see \cite{Bandos:2007wm} and refs. therein).

\section{Spinor frame approach to the 11D spinor helicity formalism }

\subsection{Scattering data in D=11}

Light--like momentum $k_{\underline{\mu} i}$ of a massless particle (consider it to be $i$-th particle of a scattering process),
 \be\label{ki2=0} k_{\underline{\mu} i}k^{\underline{\mu}}_i=0\; , \ee
 is expressed in terms of helicity spinors by
\begin{eqnarray}\label{kI=lGl}
k_{\underline{\mu} i} \delta_{\underline{q}\underline{p}}= \lambda_{\underline{\alpha} \underline{q} i}\tilde{\Gamma}{}_{\underline{\mu} }^{\underline{\alpha}\underline{\beta}}\lambda_{\underline{\beta} \underline{p} i}\; , \qquad {\Gamma}{}^{\underline{\mu} }_{\underline{\alpha}\underline{\beta}}k_{\underline{\mu} i}=
2 \lambda_{\underline{\alpha} \underline{q} i}\lambda_{\underline{\beta} \underline{q} i} \; .
 \qquad
\end{eqnarray}
Here
$$
\underline{\mu}, \underline{\nu}=0,1,...,10\; , \qquad \underline{\alpha}, \underline{\beta}=1,...,32\; , \qquad  \underline{q}, \underline{p}=1,...,16\;  \qquad
$$
and  we have used the contractions of the 11D Dirac matrices with charge conjugation matrix and its inverse, $\Gamma_{\underline \mu}{}_{\underline{\alpha}\underline \beta}:= \gamma_{\underline \mu}{}_{\underline \alpha}{}^{\underline \gamma} C{}_{\underline \gamma\beta}$ and $
\tilde{\Gamma}_{\underline \mu}{}^{\underline{\alpha}\underline \beta}:= C^{\underline{\alpha}\underline \gamma}\gamma_{\underline \mu}{}_{\underline \gamma}{}^{\underline \beta}
$, which  are real, symmetric and obey
\be\label{G=CG}
\Gamma_{\underline \mu}\tilde{\Gamma}_{\underline \nu}+ \Gamma_{\underline \nu} \tilde{\Gamma}_{\underline \mu}= \eta_{\underline \mu\underline \nu} {\bb I}_{32\times 32}\; . \qquad
\ee

Eqs. \eqref{kI=lGl} also describe the essential constraints obeyed by the helicity spinors $\lambda_{\underline{\alpha} \underline{q}}$ (denoted by $\kappa_{\frak{a} \alpha}$ in \cite{Geyer:2019ayz})  which can be solved by expressing them in terms of  spinor frame variables (spinor harmonics) \footnote{See \cite{Bandos:2007wm,Bandos:2017eof} and refs. therein for details on 11D spinor frame variables; some useful equations can be found in Appendix A of the present paper. The 11D Lorentz harmonics (which is another name for spinor moving frame variables giving credit to the ${\cal N}=2,3$ harmonic superspace approach of \cite{Galperin:1984av}) which are appropriate for the description of 11D massless superparticle were introduced for the first time in \cite{Galperin:1992pz}; the 11D harmonics appropriate for the description of 11D supermembrane were introduced and used a bit earlier in \cite{Bandos:1992hu,Bandos:1993yc}. }
\begin{eqnarray}\label{harmV=11D}
V_{\underline{\alpha}}^{(\underline{\beta})}= \left(\begin{matrix} v_{\underline{\alpha}\underline{q}}^{\; +} , & v_{\underline{\alpha} \underline{q}}^{\; -}
  \end{matrix}\right) \in Spin(1,10)\;  \qquad
 \;  \qquad
\end{eqnarray}
by \cite{Bandos:2016tsm}
\begin{eqnarray}\label{l=sqrv}
\lambda_{\underline{\alpha} \underline{q} i}=  \sqrt{\rho^\#_i} v^-_{\underline{\alpha} \underline{q} i}\; .   \qquad
\end{eqnarray}

To clarify this statement, we have to introduce a vector frame described by $SO^\uparrow (1,10)$  valued matrix
\begin{eqnarray}\label{Uab=in}
 u_{\underline{\mu}}^{(\underline{a})} = \left( {1\over 2}\left( u_{\underline{\mu}}^{=}+u_{\underline{\mu}}^{\#}
 \right), \; u_{\underline{\mu}}^{\underline{I}} \, , {1\over 2}\left( u_{\underline{\mu}}^{\#}-u_{\underline{\mu}}^{=}
 \right)\right)\; \in \; SO^\uparrow (1,10)\,   , \qquad
\end{eqnarray}
and to adapt it to our light-like momentum $k_{\underline{\mu} i}$ by assuming that one of its light-like vectors, say
$u^=_{\underline{\mu} i}= u^{0}_{\underline{\mu} i}-u^{10}_{\underline{\mu} i}$, is proportional to $k_{\underline{\mu} i}$
(see \cite{Bandos:2007wm,Bandos:2017eof} and refs. therein, in particular \cite{Sokatchev:1985tc,Sokatchev:1987nk})
\begin{eqnarray}\label{k=ru--}
k_{\underline{\mu} i} =\rho^{\#}_i u^=_{\underline{\mu} i} \; .  \qquad
\end{eqnarray}

The spinor frame variables $v^-_{\underline{\alpha} \underline{q} i}$ can be considered as a kind of square root of the light-like frame vector
$u_{\underline{\mu}}^=$ in the sense that the following constraints hold
\begin{eqnarray}\label{u--=v-v-}
 && u_{\underline{\mu}}^= \Gamma^{\underline{\mu}}_{\underline{\alpha}\underline{\beta}}= 2v_{\underline{\alpha} \underline{q}}{}^- v_{\underline{\beta} \underline{q}}{}^-  \; , \qquad
 v^-_{\underline{\alpha}\underline{q}} \tilde{\Gamma}_{\underline{\mu}}{}^{\underline{\alpha}\underline{\beta}} v^-_{\underline{\beta}\underline{p}}= u_{\underline{\mu}}^= \delta_{\underline{q}\underline{p}}. \qquad
\end{eqnarray}
This implies \eqref{kI=lGl} after \eqref{k=ru--} and \eqref{l=sqrv} are taken into account.
See Appendix A (particularly Eqs.
\eqref{u==v-v-11D}--\eqref{uIs=v+v-=11D}) for the complete set of relations between vector and spinor frame variables, \eqref{Uab=in} and \eqref{harmV=11D}. The possibility of using these and some other well known  properties of spinorial harmonics makes the understanding of spinor frame nature of the helicity spinors very useful for the work of \cite{Bandos:2016tsm,Bandos:2017zap,Bandos:2017eof} as well as for our study in this paper.

Of course, Eq. \eqref{l=sqrv} describes the real Majorana helicity spinors  for the case of momentum with positive energy, $k_0>0$, in which case also
$\rho^\#_i>0$ and $\sqrt{\rho^\#_i} $ is well defined. When describing the scattering processes one usually arranges to consider all the particles as, say outcoming, and assign a momentum with negative energy to incoming particles. Then, if $j$-th particle is incoming,  $\rho^\#_j<0$ and
one can write
$\lambda_{\underline{\alpha} \underline{q} j}=  \sqrt{|\rho^\#_j|} v^-_{\underline{\alpha} \underline{q} j}$ for real
$\lambda_{\underline{\alpha} \underline{q} j}$ and introduce the minus sign in the right hand sides of Eqs. \eqref{kI=lGl}. Alternatively, one can maintain these equations and  \eqref{l=sqrv}  as they are also for incoming particles with  $\rho^\#_j<0$, so that $\sqrt{\rho^\#_j}=i \sqrt{\rho^\#_j}$ and $\lambda_{\underline{\alpha} \underline{q} j}$ are just imaginary. We prefer this latter way of proceeding.

The helicity spinors \eqref{l=sqrv} also carry the information about polarizations of the particles, but to make it transparent we need to supply their space by an additional complex structure (see \cite{Bandos:2017zap} for the discussion). This can be encoded in the complex polarization vector. Polarization 11-vector
 $U_{\underline{\mu} i}$ of $i$-th particle (denoted by $e_\mu$ in \cite{Geyer:2019ayz})
obeys
\begin{eqnarray}\label{Uk=0}
 k_{\underline{\mu} i} U_i^{\underline{\mu}}=0\; , \qquad  U_{\underline{\mu} i} U_i^{\underline{\mu}}=0\;  \qquad
\end{eqnarray}
and can be decomposed (see \cite{Bandos:2017zap}) on the spacelike vectors of the moving frame (\ref{Uab=in}) associated to the momentum by (\ref{k=ru--}):
\begin{eqnarray}\label{U=UIuI}
U_{\underline{\mu} i}= u^{\underline{I}}_{\underline{\mu} i}U^{\underline{I}}_{i}\; , \qquad U^{\underline{I}}_{i}U^{\underline{I}}_{i}=0\; .
\;  \qquad
\end{eqnarray}
Using the constraint obeyed by vector and spinor frame variables (see \cite{Bandos:2017zap,Bandos:2017eof} and refs therein as well as \eqref{u==v-v-11D}--\eqref{uIs=v+v-=11D} in Appendix A) we find that
\begin{eqnarray}\label{UG=vUv} U\!\!\!\!/{}_{\underline{\alpha}\underline{\beta}}:= U_{\underline{\mu}}\Gamma^{\underline{\mu}}_{\underline{\alpha}\underline{\beta}}= 2
{v}^-_{(\underline{\alpha}|\underline{q}} \gamma^{\underline{I}}_{\underline{q}\underline{p}}v^+_{|\underline{\beta})\underline{p}} U^{\underline{I}}_i \; .
  \qquad
\end{eqnarray}

As it was discussed in \cite{Bandos:2017zap}, the  (complex null) polarization nine-vector $U^I$ in (\ref{U=UIuI}) can be related  by
\begin{eqnarray}\label{Ug=bwbw} U\!\!\!\!/{}_{\underline{q}\underline{p}i}:= U^{\underline{I}}_i \gamma^{\underline{I}}_{\underline{q}\underline{p}}= 2 \bar{w}_{\underline{q}\underline{A} i}\bar{w}_{\underline{p}\underline{A}i }\; \qquad
\end{eqnarray}
to the complex  $16\times 8$ matrices obeying 'purity conditions' (in terminology of \cite{Geyer:2019ayz})
\begin{eqnarray}\label{bwbw=0}
\bar{w}_{{\underline{q}}\underline{A}} \bar{w}_{{\underline{q}}\underline{B}} =0\; , \qquad \; \underline{A}, \underline{B} =1,...,8\;  \qquad
\end{eqnarray}
(for shortness, here and below  we omit the index $i$ enumerated scattering particles when this cannot lead to a confusion).

Actually,  $\bar{w}_{{\underline{q}}\underline{A}}$ are
internal frame variables \cite{Bandos:2017zap} or $SO(9)/SO(7)\times SO(2)$  harmonics   (in the sense of  \cite{Galperin:1984av}, see \cite{Bandos:2017zap} and refs therein). This is to say they are 8 complex linear combinations of columns of an $SO(9)$ valued matrix, schematically
\begin{eqnarray}\label{harm=SO9}
(\bar{w}_{\underline{q}\underline{A}}, {w}_{\underline{q}}{}^{\underline{A}})
\;  \in \; SO(9)\;  \qquad
\end{eqnarray}
with ${w}_{\underline{q}}{}^{\underline{A}}=(\bar{w}_{\underline{q}\underline{A}})^*$, defined up to $SO(7)\times SO(2)$ gauge transformations. Eq. \eqref{harm=SO9} implies
that $\bar{w}_{\underline{q}\underline{A}}$ and $ {w}_{\underline{q}}{}^{\underline{A}}$ obey
\begin{eqnarray}\label{wbw+cc=1}
&& {w}_{\underline{q}}{}^{\underline{A}} \bar{w}_{\underline{p}\underline{A}}+ \bar{w}_{\underline{q}\underline{A}}{w}_{\underline{p}}{}^{\underline{A}} =\delta_{\underline{q}\underline{p}}\; , \qquad \\
\label{bww=1}
&& \bar{w}_{\underline{q}\underline{B}}w_{\underline{q}}{}^{\underline{A}}=\delta_{\underline{B}}{}^{\underline{A}}\; , \qquad w_{\underline{q}}{}^{\underline{A}} w_{\underline{q}}{}^{\underline{B}} =0 \; , \qquad \bar{w}_{\underline{q}\underline{A}} \bar{w}_{\underline{q}\underline{B}} =0\;   \qquad
\end{eqnarray}
(the set of which includes \eqref{bwbw=0}) as well as  \eqref{Ug=bwbw} and a few similar relations with other vectors of $SO(9)$ vector frame which can be found in \cite{Bandos:2017zap} and in Appendix A.2
\footnote{\label{foot-SO16} In 10D case the counterparts of Eqs. \eqref{wbw+cc=1} and \eqref{bww=1} with
$\underline{q}\mapsto q=1,...,8$ and $\underline{A}\mapsto A=1,...,4$
guarantee that the matrix $(\bar{w}_{{q}{A}}, {w}_{{q}}{}^{{A}})
\;  \in \; SO(8)$. In our 11D case Eqs. \eqref{wbw+cc=1} and \eqref{bww=1} imply only $(\bar{w}_{\underline{q}\underline{A}}, {w}_{\underline{q}}{}^{\underline{A}})
\;  \in \; SO(16)$ while the reduction to $SO(9)$ is achieved  by imposing additional relations \eqref{bwgIbw=UI}, \eqref{wgIw=bUI} and  \eqref{bwgIbw=UIK} }.

\subsection{Complex spinor frame variables and complex helicity spinors}

As in \cite{Bandos:2017zap}, it will be convenient to introduce the set of {\it complex spinor harmonics} (complex spinor frame variables)
composed of the real spinor frame variables \eqref{harmV=11D} and internal harmonics \eqref{harm=SO9}  according to
\begin{eqnarray}\label{v-A:==11D}
 v_{\underline{\alpha} \underline{A}}^{-}:= v_{\underline{\alpha} \underline{q}}^{-} \bar{w}_{\underline{q}\underline{A}}\; , \qquad \bar{v}{}_{\underline{\alpha}}^{-\underline{A}}:= v_{\underline{\alpha} \underline{q}}^{-} {w}_{\underline{q}}^{\; \underline{A}} \; , \qquad v_{\underline{\alpha} \underline{A}}^{+}:= v_{\underline{\alpha} \underline{q}}^{+} \bar{w}_{\underline{q}\underline{A}}\; , \qquad  \bar{v}{}_{\underline{\alpha}}^{+\underline{A}}:= v_{\underline{\alpha} \underline{q}}^{+} {w}_{\underline{q}}^{\; \underline{A}} \;  . \qquad
\end{eqnarray}
By construction,
\begin{eqnarray}\label{v-Av-B=0} {v}^-_{\underline{\alpha}\underline{A}}{v}^{-\underline{\alpha}}_{\underline{B}}=0 \; , \qquad  {v}^-_{\underline{\alpha}\underline{A}}\bar{v}^{-\underline{\alpha}\underline{B}}=0 \; , \qquad   {v}^+_{\underline{\alpha}\underline{A}}{v}^{-\underline{\alpha}}_{\underline{B}}=0\; , \qquad {v}^+_{\underline{\alpha}\underline{A}}\bar{v}^{-\underline{\alpha}\underline{B}}=\delta_{\underline{A}}{}^{\underline{B}}\; , \qquad \ldots \; . \qquad
\end{eqnarray}
With this notation, Eqs. (\ref{UG=vUv}), (\ref{Ug=bwbw})  imply
\begin{eqnarray}\label{UG=vv} U\!\!\!\!/{}_{\underline{\alpha}\underline{\beta}}:= U_{\underline{\mu}} \Gamma^{\underline{\mu}} _{\underline{\alpha}\underline{\beta}}=
4
{v}^-_{(\underline{\alpha}|\underline{A}}v^+_{|\underline{\beta})\underline{A}}
\;  . \qquad
\end{eqnarray}

Below we find convenient to use the $SO(1,1)$ invariant complex helicity spinors
\begin{eqnarray}\label{lA:==11D}
\lambda_{\underline{\alpha} \underline{A}}= \sqrt{\rho^{\#}}v_{\underline{\alpha} \underline{A}}^{-}
 \; , \qquad \bar{\lambda}{}_{\underline{\alpha}}^{\; \underline{A}}=\sqrt{\rho^{\#}} \bar{v}{}_{\underline{\alpha}}^{-\underline{A}}  \;  \qquad
\end{eqnarray}
instead of $v_{\underline{\alpha} \underline{A}}^{-}$ and $ \bar{v}{}_{\underline{\alpha}}^{-\underline{A}}$ so that the second equation in (\ref{kI=lGl}) can be written in an equivalent form
\begin{eqnarray}\label{kG=rvAvA}
\tilde{k}\!\!\!/{}^{\underline{\alpha}\underline{\beta}}= 4\rho^{\#}
{v}^{-(\underline{\alpha}}_{\underline{A}}\bar{v}{}^{-\underline{\beta})\underline{A}}= 4
\lambda^{(\underline{\alpha}}_{\underline{A}}\bar{\lambda}{}^{\underline{\beta})\underline{A}}\;\qquad \Leftrightarrow \qquad  {k}\!\!\!/{}_{\underline{\alpha}\underline{\beta}}= 4\rho^{\#}
\bar{v}{}_{(\underline{\alpha}}^{- \underline{A}} v_{\underline{\beta})\underline{A}}^{\;\;\;  -}=4
\bar{\lambda}{}_{(\underline{\alpha}}^{\;\underline{A}} \lambda_{\underline{\beta})\underline{A}}\;  . \qquad
\end{eqnarray}
However, we do not find practical to introduce also $SO(1,1)$  invariant counterparts of the complementary  spinors $ v_{\underline{\alpha} \underline{A}}^{+}$ and $ \bar{v}{}_{\underline{\alpha}}^{+\underline{A}}$ from the spinor frame.
Of course, if we wish to present e.g.  Eq.  \eqref{UG=vv} literally but in terms of  the helicity spinors, we obtain not so elegant $U\!\!\!\!/{}_{\underline{\alpha}\underline{\beta}}=
4\lambda_{(\underline{\alpha}|\underline{A}}v^+_{|\underline{\beta})\underline{A}}/\sqrt{\rho^{\#}}$. However, instead we can   write the following equivalent set of relations involving $U\!\!\!\!/{}_{\underline{\alpha}\underline{\beta}}$ and the constrained spinors \eqref{lA:==11D} only:
\bea\label{UluA=ldA}
\tilde{U}\!\!\!\!/{}^{\underline{\alpha}\underline{\beta}}_i \lambda_{\underline{\beta}\underline{A}i}=0\; , \qquad
\tilde{U}\!\!\!\!/{}^{\underline{\alpha}\underline{\beta}}_i \bar{\lambda}{}_{\underline{\beta}i}{}^{\underline{A} }=-2 \lambda_{\underline{A}i}{}^{\underline{\alpha}}\; .
\eea

Using \eqref{v-Av-B=0} it is not difficult to check  that
\bea\label{klA=0}
\tilde{k}\!\!\!/{}^{\underline{\alpha}\underline{\beta}}_i \lambda_{\underline{\beta}\underline{A}i}=0\; , \qquad
\tilde{k}\!\!\!/{}^{\underline{\alpha}\underline{\beta}}_i\bar{\lambda}{}_{\underline{\beta}i}^{\underline{A} }=0\; .
\eea
The first equations in \eqref{UluA=ldA} and  \eqref{klA=0} together with simple counting arguments
imply that $\lambda_{\underline{\alpha} \underline{A} i}$ (or  $ v_{\underline{\alpha} \underline{A} i}^{-}$) provide a basis for the common  kernel space of
$ k\!\!\!/{}_{\underline{\alpha}\underline{\beta}i}$ and $ U\!\!\!\!/{}_{\underline{\alpha}\underline{\beta}i}\quad $ \footnote{The elements of this basis, $\lambda _{\underline{\alpha} \underline{A} i}$,  were denoted by $\epsilon_{\frak{a}a}=\kappa_{\frak{a}\alpha}\epsilon_{\alpha a}$ in \cite{Geyer:2019ayz} where
$\epsilon_{\alpha a}$ is the notation for $ \bar{w}_{\underline{q}\underline{A}}$.
}
\begin{eqnarray}\label{kGchi=0}
 \tilde{k}\!\!\!/{}^{\underline{\alpha}\underline{\beta}}_i \chi_{\underline{\beta}i}=0= \tilde{U}\!\!\!\!/{}^{\underline{\alpha}\underline{\beta}}_i \chi_{\underline{\beta}i}\qquad \Rightarrow \qquad \chi_{\underline{\beta}i}= \chi^{A} \lambda_{\underline{\beta} \underline{A}i}\equiv  \chi^{+A} v^-_{\underline{\beta} \underline{A}i}
\; .  \qquad
\end{eqnarray}

Then the second equations in \eqref{klA=0} and \eqref{UluA=ldA}  indicate that the  set of constrained spinors $\bar{\lambda}{}_{\underline{\alpha}i}^{\; \underline{A}}$ complete
$\lambda_{\underline{\alpha} \underline{A} i}$ till the basis of the space of solutions of the massless Dirac  equation,
while the matrix  $ \tilde{U}\!\!\!\!/{}^{\underline{\alpha}\underline{\beta}}$ maps these into $\lambda_{\underline{\alpha} \underline{A} i}$,
\bea\label{UlA=lA}
\tilde{k}\!\!\!/{}^{\underline{\alpha}\underline{\beta}}_i \lambda_{\underline{\beta}i}{}^{\underline{A}}=0\; , \qquad
U\!\!\!\!/{}_{\underline{\alpha}\underline{\beta}i} \bar{\lambda}{}^{\underline{\beta}\underline{A}}{}_i= -2\lambda_{\underline{\alpha} \underline{A} i}\; .
\eea
This allows us to state that ${\lambda}{}_{\underline{\alpha}i}^{\; \underline{A}}$  provide the basis of the complementary to the space of common zero modes of
$ k\!\!\!/{}_{\underline{\alpha}\underline{\beta}i}$ and $ U\!\!\!\!/{}_{\underline{\alpha}\underline{\beta}i}$ in the space of the  solutions of 11D massless Dirac equation.

With Eqs. \eqref{v-Av-B=0}  we also find
\begin{eqnarray}\label{lAlB=0}
\lambda_{\underline{\alpha}\underline{A}}\lambda_{\underline{B}}{}^{\underline{\alpha}}&=&0 \; , \\ \label{v-Av-B=G2kU}
{\lambda}_{\underline{\alpha}\underline{A}}{\lambda}_{\underline{A}}{}^{\underline{\beta}}&:=&\rho^\# {v}^-_{\underline{\alpha}\underline{A}}{v}^{-\underline{\beta}}_{\underline{A}}= - \frac 1 4 k_\mu U_\nu \Gamma^{\mu\nu}{}_{\underline{\alpha}}{}^{\underline{\beta}}\;
\end{eqnarray}
and
\begin{eqnarray}\label{v-AG2v-B=kU}
\lambda_{\underline{A}}\Gamma_{\underline{\mu}\underline{\nu}}\lambda_{\underline{B}}=\rho^{\#} v^{-}_{\underline{A}}\Gamma_{\underline{\mu}\underline{\nu}}v^{-}_{\underline{B}}= + k_{[\mu} U_{\nu]}\delta_{\underline{A}\underline{B}} \; .
\end{eqnarray}
One can recognize in (\ref{v-Av-B=G2kU}) and \eqref{v-AG2v-B=kU} the relations from (2.5) of \cite{Geyer:2019ayz}.
Our spinor frame approach is very efficient in derivation of such type
relations.

Notice that the indices of, say,
$ \lambda_{\underline{\alpha} \underline{A} i}$ and $ v_{\underline{\alpha} \underline{A} i}^{-}$ are transformed by the rigid $Spin(1,10)$ group, common for all
values of $i$,  and by $Spin(7)_i$ transformations, specific for each of the scattered particles.
The internal harmonics $ \bar{w}_{\underline{q}\underline{A} i}$ are transformed by $Spin(9)_i\otimes Spin (7)_i$, where $Spin(9)_i$ is also specific for $i$-th particle.

\section{Polarized scattering equation of 11D supergravity}

\subsection{Scattering equations}

Scattering equations \cite{Gross:1987kza,Gross:1987ar,Cachazo:2013hca,Cachazo:2014xea}
establishing the relation between scattered particles and points $\sigma_i$ on Riemann sphere read
\begin{eqnarray}\label{ScEq1}
 \sum_{j=1}^n \frac {k^{\mu}_ik_{j\mu}} {\sigma_i-\sigma_j}=0\; .  \qquad
\end{eqnarray}
In this subsection we omit underlining of 11D indices to stress that the equations are valued for arbitrary $D$.

As in \cite{Geyer:2019ayz} (see also refs. therein) we can introduce the meromorphic D-vector function
\begin{eqnarray}\label{Pmu=}
P_\mu (\sigma) = \sum_i \frac {k_{i\mu}} {\sigma-\sigma_i}\;   \qquad
\end{eqnarray}
and to write the scattering equation (\ref{ScEq1}) in the form
\begin{eqnarray}\label{kiPsi=0}
k^{\mu}_i P_\mu (\sigma_i) = \sum_{j\not= i} \frac {k^{\mu}_ik_{j\mu}} {\sigma_i-\sigma_j}=0\; .  \qquad
\end{eqnarray}
Notice that, while $P_\mu (\sigma_i)$ diverges, its contraction with $k^{\mu}_i$ is well  defined
(if no one of $\sigma_{j\not= i}$ coincide with  $\sigma_{i}$, as usually assumed) due to the mass-shell conditions \eqref{ki2=0}.

One can also write the scattering equation (\ref{ScEq1}) as equation on the meromorphic vector function (\ref{Pmu=}) only:
\begin{eqnarray}\label{ResP2=0=}
Res_{\sigma=\sigma_i} \frac 1 2 P^2(\sigma)=0\;  . \qquad
\end{eqnarray}
This equation  actually implies (see e.g. \cite{Geyer:2019ayz} and refs. therein) the light-likeness of the meromorphic D--vector function (\ref{Pmu=}),
\begin{eqnarray}\label{P2=0}
P^\mu (\sigma) P_\mu  (\sigma) = 0\;   \qquad
\end{eqnarray}
for any $\sigma$. Thus we consider (\ref{P2=0}) with (\ref{Pmu=})  as the third equivalent form of the scattering equation.

The constraint (\ref{P2=0}) can be generated from the so-called ambitwistor string action \cite{Mason:2013sva} and Eq. (\ref{Pmu=}) can be obtained from the deformation of this action obtained by incorporating the contribution  of the suitable vertex operators to the path integral measure. Below we will describe  11D supersymmetric generalization of the ambitwistor superstring action proposed in \cite{Bandos:2014lja} (see \cite{Bandos:2006af,Bandos:2007qn} for earlier discussion in the context of twistor string). In
\cite{Geyer:2019ayz} a modified version of this action is discussed; this paper gives the  arguments in favour of the original action.

\subsection{Constrained spinor function on Riemann sphere }
\label{sec=lambda}

Eq. (\ref{P2=0}) suggests the existence of a meromorphic function carrying 11D spinor index which plays the role of square root of the above meromorphic vector function in the same sence as helicity spinors can be associated with square roots of the light-like momentum, (\ref{kI=lGl}),
\begin{eqnarray}\label{Pmu=lGl}
P_{\underline{\mu}}(\sigma) \delta_{\underline{q}\underline{p}}= \lambda_{\underline{q}}(\sigma)\tilde{\Gamma}{}_{\underline{\mu} }\lambda_{\underline{p}}(\sigma)\; , \qquad
2 \lambda_{\underline{\alpha} \underline{q}}(\sigma) \lambda_{\underline{\beta} \underline{q}}(\sigma)= {\Gamma}{}^{\underline{\mu} }_{\underline{\alpha}\underline{\beta}}P_{\underline{\mu}}(\sigma)\; .
 \qquad
\end{eqnarray}
Furthermore, then it is convenient to introduce a spinor frame field $ v^-_{\underline{\alpha} \underline{q}}(\sigma )$ \cite{Bandos:2014lja} and a (purely gauge or St\"uckelberg) density $\rho^\#(\sigma )$ and to use this to write the general solution of the constraints \eqref{Pmu=lGl} in the form
\begin{eqnarray}\label{l=r12v}
\lambda_{\underline{\alpha} \underline{q}}(\sigma )=  \sqrt{\rho^\#(\sigma )} v^-_{\underline{\alpha} \underline{p}}(\sigma ) {\cal S}_{\underline{p}\underline{q}} (\sigma )
\; ,
 \qquad {\cal S}_{\underline{p}\underline{r}} {\cal S}_{\underline{q}\underline{r}} = {\delta}_{\underline{p}\underline{q}} \; . \qquad
\end{eqnarray}

Indeed, substituting \eqref{l=r12v} into \eqref{Pmu=lGl} we find
\begin{eqnarray}\label{P=lGl}
P_{\underline{\mu}}(\sigma ) \delta_{\underline{q}\underline{p}}= \rho^{\#}(\sigma)  v^-_{ \underline{q}}(\sigma )\tilde{\Gamma}{}_{\underline{\mu} }v^-_{\underline{p}}(\sigma )\; , \qquad
2 \rho^{\#}(\sigma) v^-_{\underline{\alpha} \underline{q}}(\sigma)v^-_{\underline{\beta} \underline{q}}(\sigma)= {\Gamma}{}^{\underline{\mu} }_{\underline{\alpha}\underline{\beta}}P_{\underline{\mu}}(\sigma)\; .
 \qquad
\end{eqnarray}
which describe  the essential constraints on the spinor frame functions and their relation with the meromorphic vector function,
\be\label{P=ru=} P_{\underline{\mu}}(\sigma ) =\rho^{\#}(\sigma) u^=_{\underline{\mu}}(\sigma )\; . \qquad \ee

Notice that the algebraic relations between spinor functions, spinor frame fields and the meromorphic vector function obeying \eqref{P2=0} are the same as \eqref{kI=lGl}, \eqref{k=ru--}, \eqref{u--=v-v-} relating the helicity spinors, spinor frame variables and light--like momentum of $i$-th scattered particle. This is why we use essentially the same symbols in both cases (distinguishing them by indicating explicitly the dependence on $\sigma$ in the case of functions and putting the index $i$   in the case of variables corresponding to $i$-th scattered particle).

The presence of $SO(16)$ valued matrix field ${\cal S}(\sigma )\in SO(16)$ (${\cal S}{\cal S}^T=I$) in \eqref{l=r12v}
reflects the invariance of \eqref{Pmu=lGl} under the $SO(16)$ gauge transformations. One might wonder why we have not introduced such a matrix in the relation \eqref{l=sqrv} between polarization spinors corresponding to $i$-th of scattering particles and the $i$-th spinor frame. The reason is that the helicity spinors should also carry the information about particle polarizations. This is encoded in the polarization vector which is represented by complex SO(9) vector with vanishing square. Its relation with the complex helicity spinors  described by Eq. \eqref{UG=vv} requires the identification of the 16 component index $q$ of the real helicity spinor as $SO(9)$ spinor index thus  breaking $SO(16)$ symmetry of   Eqs. \eqref{kI=lGl} down to $SO(9)$ and prohibiting  the inclusion of $SO(16)$ matrix in the common solution \eqref{l=sqrv} of   \eqref{kI=lGl} and \eqref{UG=vv}. In contrast, the spinorial functions should obey, at present stage, only the constraints \eqref{Pmu=lGl} which are invariant under $SO(16)$ gauge symmetry, so that its general solution is given by \eqref{l=r12v}.


The meromorphic spinor function
$\lambda_{\underline{\alpha} \underline{q}}(\sigma )$ which would correspond to the vector meromorphic function of Eq. (\ref{Pmu=}) in the sense of Eqs. (\ref{Pmu=lGl}) should have the structure similar to (\ref{Pmu=}), but with the use of helicity spinors (or spinor frame variables) related to light-like momenta by (\ref{kG=rvAvA}) instead of the momenta itself.
The expression of such a type was proposed in \cite{Geyer:2019ayz}. However, the moving frame treatment of the 11D helicity spinors makes manifest that this was the gauge fixing description.

The complete gauge covariant form of such relation reads
\begin{eqnarray}\label{sol=pScEq1} \lambda_{\underline{\alpha} \underline{q}}(\sigma )= \sum_{i=1}^n \sqrt{\rho^\#_i}\, \frac { v_{\underline{\alpha}\underline{A}i}^{\; -}{W}_{\underline{q} i}^{\; \underline{A}}(\sigma) }{\sigma-\sigma_i} = \sum_{i=1}^n \, \frac { \lambda_{\underline{\alpha}\underline{A}i}{W}_{\underline{q} i}^{\; \underline{A}}(\sigma) }{\sigma-\sigma_i}
\; ,
\qquad
\end{eqnarray}
where the function ${W}_{\underline{q} i}^{\; \underline{A}}(\sigma)$ has no poles and obeys
 the 'purity' conditions
 \begin{eqnarray}\label{WqAWqB=0}
{W}_{\underline{q}i}^{\underline{A}}(\sigma) {W}_{\underline{q}i}^{\underline{B}} (\sigma)=0 \; .
\end{eqnarray}
This  is necessary to obey the constraints (\ref{Pmu=lGl}) with  meromorphic 11-vector (\ref{Pmu=}).
Indeed, taking into account  (\ref{sol=pScEq1}),  (\ref{Pmu=}) and (\ref{kG=rvAvA}), we can write  Eq. (\ref{Pmu=lGl}) in the form
\begin{eqnarray}\label{ll=PG=}
\sum\limits_i \frac {\lambda_{\underline{\alpha}\underline{A}i }}{\sigma-\sigma_i}  \sum\limits_{j} \frac {\lambda_{\underline{\beta}  \underline{B} j}}{\sigma-\sigma_j} W_{\underline{q}j}^{\underline{B}}(\sigma) W_{\underline{q}i}^{\underline{A}}(\sigma)= \sum\limits_i \frac {2\lambda_{(\underline{\alpha} | \underline{A}i }^{\;\; }  \lambda_{|\underline{\beta} ) i}^{\, \underline{A} }} {\sigma-\sigma_i}\; .
\qquad
\end{eqnarray}
When all $\sigma_i$'s are different, the r.h.s. of this equation has first order poles at $\sigma=\sigma_i$ with residues
$2\lambda_{(\underline{\alpha} | \underline{A}i }  \lambda_{|\underline{\beta} ) i}^{\, \underline{A} }= 2{\rho^{\#}_i} v_{(\underline{\alpha} | \underline{A}i }^{\;\; -}  v_{|\underline{\beta} ) i}^{\, \underline{A} -}\equiv {\rho^{\#}_i} v_{\underline{\alpha}  \underline{q}i }^{\;\; -}  v_{\underline{\beta}  \underline{q}i}^{\;\; -}$. In contrast, the l.h.s generically has second order poles. These vanish if we require
 ${W}_{\underline{q}i}^{\underline{A}} (\sigma)$ to obey the 'purity' conditions \eqref{WqAWqB=0}.

Notice that the r.h.s. of Eq. \eqref{sol=pScEq1} is clearly complex, as meromorphic function should be, so that our spinor functions
$\lambda_{\underline{\alpha} \underline{q}}(\sigma )$ are not real. Such a complexification is characteristic for the ambitwistor string and CHY scattering equation approaches, as well as {\it e.g.} for the pure spinor description of quantum 10D superstrings \cite{Berkovits:2000fe,Berkovits:2004px,Berkovits:2006vi,Berkovits:2017ldz}. Already the form of the vector function \eqref{Pmu=} indicates that it is complex and hance complex are its square roots in the sense of Eqs. \eqref{Pmu=lGl} and \eqref{P=lGl}. Thus also the spinor moving frame field
${v}_{\underline{\alpha}\underline{q}i}^{\; -} (\sigma)$, moving frame field ${u}_{\underline{\mu}}^{=} (\sigma)$ and density $\rho^\#(\sigma)$ in \eqref{l=r12v}, \eqref{P=lGl} and \eqref{P=ru=} are  complexification of the functions used e.g. in \cite{Bandos:2007wm} and \cite{Bandos:2006af}. The matrix ${\cal S}$ in \eqref{l=r12v} should be also considered as complex so that, strictly speaking,  it takes values in $SO(16, {\bb C})$.
As far as the counting of degrees of freedom is concerned, the usual strategy in the models with complexified variables is to substitute reality by analyticity, {\it i.e.} to allow for the dependence on, say, complex $\lambda_{\underline{\alpha} \underline{q}}(\sigma )$ but not on its complex conjugate.

\subsection{Preliminaries on $SO(16)$ gauge symmetry, its naturalness and St\"uckelberg realization}

The appearance of the matrix functions  ${W}_{\underline{q}i}^{\underline{A}} (\sigma)$ and not just constant matrix
in the r.h.s of  \eqref{sol=pScEq1} is necessary to make equations gauge invariant. To motivate the  requirement of gauge
invariance we can  turn to the ambitwistor superstring origin of the spinorial function $\lambda_{\underline{\alpha} \underline{q}}(\sigma )$ providing the square root of the meromorphic vector function \eqref{Pmu=} in the sense of \eqref{Pmu=lGl}.

Even in the case if the relations of constrained spinor functions
$\lambda_{\underline{\alpha} \underline{q}}(\sigma )$ with spinor frame field in \eqref{l=r12v}  were not including the
$SO(16)$ matrix field and were just
$\lambda_{\underline{\alpha} \underline{q}}(\sigma )=  \sqrt{\rho^\#(\sigma )} v^-_{\underline{\alpha} \underline{q}}(\sigma )$
(the counterpart of this situation we will observe in 10D case), the r.h.s. of \eqref{sol=pScEq1} should include the matrix {\it field} anyway. This is because the spinor frame field  $v^-_{\underline{\alpha} \underline{q}}(\sigma )$  suitable for the description of 11D ambitwistor string (and tensionless superstring) is defined up to $SO(9)$ gauge symmetry transformations with $\sigma$-dependent parameters which should act also on ${W}_{\underline{q}i}^{\underline{A}} (\sigma)$ to leave Eq. \eqref{sol=pScEq1} gauge invariant.

In D=11 the relation of the spinor function  and spinor frame functions \eqref{l=r12v} includes $SO(16)$ valued matrix  ${\cal S}\in SO(16)$, so that the reference on defining gauge symmetry of the spinor moving frame field is not valid and the arguments should be different. A way which is more straightforward, although probably not so  convincing by itself, consists in just stating  that the matrix field ${\cal S}(\sigma)$ should not carry additional degrees of freedom which can be provided by   imposing the requirement of
$SO(16)$ gauge symmetry  acting  on $\lambda_{\underline{\alpha} \underline{q}}(\sigma )$ as
 \begin{eqnarray}\label{l-lO}
\lambda_{\underline{\alpha} \underline{q}}(\sigma )\mapsto  \lambda_{\underline{\alpha} \underline{p}}(\sigma ) {\cal O}_{\underline{p} \underline{q}}(\sigma )
\;  \qquad with \qquad {\cal O}(\sigma ){\cal O}^T(\sigma )={\bb I}\;
 \qquad
\end{eqnarray}
and  leaving invariant \eqref{l=r12v}.
The real argument in favour of this requirement is that, as we will see below, $SO(16)$  is also a gauge symmetry of the 11D ambitwistor superstring action in its supertwistor form.

To leave  invariant Eq. (\ref{sol=pScEq1}), this gauge symmetry
should also act on the matrix function ${W}_{\underline{q} i}^{\; \underline{A}}(\sigma)$,
 \begin{eqnarray}\label{W-WU}
{W}_{\underline{q} i}^{\; \underline{A}}(\sigma)\mapsto  {W}_{\underline{p} i}^{\; \underline{A}}(\sigma) {\cal O}_{\underline{p} \underline{q}}(\sigma )
\; .
 \qquad
\end{eqnarray}
Thus the requirement of $SO(16)$ gauge covariance do not allow us to write a constant matrix ${W}_{\underline{q} i}^{\; \underline{A}}$ in the r.h.s of Eq. (\ref{sol=pScEq1}), as it was written in its counterpart presented in \cite{Geyer:2019ayz}.
On the other hand, as we are going to show, after imposing on ${W}_{\underline{q} i}^{\; \underline{A}} (\sigma)$ some additional conditions, one can fix  a gauge with respect to the $SO(16)$ gauge symmetry in which
${W}_{\underline{q} i}^{\; \underline{A}} (\sigma)$ for a given $i$ coincides with some ${W}_{\underline{q} i}^{\; \underline{A}}$. This  implies
\begin{eqnarray}\label{Wi=WiOi}{W}_{\underline{q} i}^{\; \underline{A}} (\sigma)= {W}_{\underline{p} i}^{\; \underline{A}}\,  \tilde{\cal O}_{i\underline{p} \underline{q}}(\sigma )\; . \qquad
\end{eqnarray}

Furthermore, in sec. \ref{Ambi-Pol}  we will derive Eq. (\ref{sol=pScEq1}) from 11D ambitwistor superstring model and  show that  the stronger version of Eq. \eqref{Wi=WiOi}, which includes the same $SO(16)$ valued matrix field
$\tilde{\cal O}_{i\underline{p} \underline{q}}(\sigma )=\tilde{\cal O}_{\underline{p} \underline{q}}(\sigma ) $ for all values of  $i$, holds:
\begin{eqnarray}\label{Wi=WiO}
{W}_{\underline{q} i}^{\; \underline{A}} (\sigma)= {W}_{\underline{p} i}^{\; \underline{A}}\,  \tilde{\cal O}_{\underline{p} \underline{q}}(\sigma ) \; . \qquad
\end{eqnarray}
This makes manifest the existence of the gauge in which the  expression similar to the one proposed in  \cite{Geyer:2019ayz} appears
\footnote{A derivation  of the gauge fixed expression was discussed schematically in  \cite{Geyer:2019ayz}, but a number of issues were obscure in this discussion. Here we will present a clean derivation which requires, in particular,  the use of an embedding of the 11D ambitwistor superstring model into an enlarged superspace.}.

On the other hand, \eqref{Wi=WiO} implies that this $SO(16)$ is realized as a St\"uckelberg gauge symmetry. The reason for this will be clarified below. What happens is that, while the $SO(16)$ is a true gauge symmetry of the ambitwistor superstring action, which is originally hidden but can be made manifest in its  supertwistor formulation, it is broken by the vertex operators of physical states. To preserve it in the ambitwistor superstring action deformed by a term accounting for the contribution of the vertex operator to the path integral, $SO(16)$ valued St\"uckelberg field, $\tilde{\cal O}_{\underline{p} \underline{q}}(\sigma )\in SO(16)$ in \eqref{Wi=WiO}, must be introduced.

\subsection{Polarized scattering equation}

Now let us observe that the residues of the poles of l.h.s. and r.h.s. of Eq.  (\ref{ll=PG=})   coincide if
the spinor frames and polarization data associated to the scattered particles are related by the condition
\begin{eqnarray}\label{ScEqvi=vj}
\sum_{j}\sqrt{\rho^\#_j} \frac {v_{\underline{\alpha} \underline{B}j}^- W_{\underline{q}j}^{\underline{B}}W_{\underline{q}i}^{\underline{A}}}{\sigma_i-\sigma_j}=\sqrt{\rho^\#_i}\bar{v}{}_{\alpha i}^{-\underline{A}}\;
\end{eqnarray}
or
\begin{eqnarray}\label{ScEqli=lj}
\sum_{j}\frac {\lambda_{\underline{\alpha} \underline{B}j} W_{\underline{q}j}^{\underline{B}}W_{\underline{q}i}^{\underline{A}}}{\sigma_i-\sigma_j}=\bar{\lambda}_{\alpha i}{}^{\underline{A}}\; .
\end{eqnarray}
Using \eqref{sol=pScEq1} we can write this equation  in a bit more compact equivalent form
\begin{eqnarray}\label{pScEq}
\lambda_{\underline{\alpha} \underline{q}}(\sigma_i ) {W}_{\underline{q}i}^{\underline{A}}(\sigma_i ) =   \sqrt{\rho^\#_i} \bar{v}_{\underline{\alpha}i}^{-\underline{A}}=\bar{\lambda}_{\alpha i}{}^{\underline{A}} \; . \qquad
\end{eqnarray}
This relation basically coincides with the one  first introduced in \cite{Geyer:2019ayz} and called there {\it 11D polarized scattering equation}.
Our study revealed the moving frame nature  of both the constrained spinors and constrained spinor functions involved in it. Furthermore, the difference with \cite{Geyer:2019ayz} is that the l.h.s of our version of the polarized scattering equation includes  a value of a(n analytic) matrix function \eqref{Wi=WiO} at $\sigma=\sigma_i$,
${W}_{\underline{q}i}^{\underline{A}}(\sigma_i )={W}_{\underline{p}i}{}^{\underline{A}}\,\tilde{{\cal O}}_{\underline{p}\underline{q}}(\sigma_i)$, rather than just a constant matrix ${W}_{\underline{q}i}{}^{\underline{A}}$.
The reason for this is that in such a way we make the polarized scattering equation invariant under the SO(16) gauge symmetry characteristic, as we will see below, for ambitwistor superstring. Furthermore,  just our $SO(16)$ covariant version of the expression for the meromorphic spinor function \eqref{sol=pScEq1} can be obtained naturally from the  ambitwistor superstring action deformed by an appropriate vertex operator contribution.

Eq. \eqref{ScEqli=lj} also can (or rather must) be called polarized scattering equation. This is a 'polarized' counterpart of the scattering equation (\ref{ScEq1}) while (\ref{pScEq})  is a polarized counterpart of the scattering equation in its form of Eq. (\ref{kiPsi=0}).

When  obtaining \eqref{pScEq} from \eqref{ScEqli=lj} we have used the fact that, as a consequence of
\eqref{Wi=WiO},
\begin{eqnarray}\label{WW=WsWs}
W_{\underline{q}j}^{\underline{B}}(\sigma)W_{\underline{q}i}^{\underline{A}}(\sigma)=
W_{\underline{q}j}^{\underline{B}}(\sigma_i)W_{\underline{q}i}^{\underline{A}}(\sigma_i)=
W_{\underline{q}j}^{\underline{B}}W_{\underline{q}i}^{\underline{A}}\; .
\end{eqnarray}
Thus the presence of constant matrices $W_{\underline{q}i}^{\underline{A}}$ in  \eqref{ScEqli=lj} does not contradict the statement of   $SO(16)$ gauge invariance of the polarized scattering equation \eqref{pScEq}.

It is not difficult to observe  that $j=i$ contribution to the l.h.s. of Eq. \eqref{ScEqli=lj}, which might produce a singularity, vanishes due to the 'purity' conditions (\ref{WqAWqB=0}), so that an equivalent form of that equation is
\begin{eqnarray}\label{ScEqvi=vj1}
\sum_{j\not=i} \frac {\lambda_{\underline{\alpha} \underline{B}j} W_{\underline{q}j}^{\underline{B}}W_{\underline{q}i}^{\underline{A}}}{\sigma_i-\sigma_j}=\bar{\lambda}_{\underline{\alpha} i}{}^{\;\underline{A}}\; . \qquad
\end{eqnarray}

The polarized scattering equation is expected to be a condition on the scattering data: momenta and polarizations of the scattered particle. Then $W_{\underline{q}i}^A $ entering \eqref{ScEqvi=vj1} should describe the  data related to $i$-th of the scattered particle. This suggests to identify it with  the internal frame matrix variable $w_{\underline{q}i}^A$ \eqref{harm=SO9}
 \begin{eqnarray}\label{Wi=wi}
 W_{\underline{q}i}^A = w_{\underline{p}i}^A \; . \qquad
\end{eqnarray}
We however, restrain ourselves from fixing this identification rigidly at this stage of development of the formalism and, keeping in mind \eqref{Wi=wi}, keep below a separate notation $W_{\underline{q}i}^A $ for the matrix entering the scattering equation.

Resuming, the polarized scattering equation  (\ref{ScEqli=lj}) guarantees that $\lambda_{\underline{\alpha} \underline{q}}(\sigma)$ of (\ref{sol=pScEq1})  obeys $Res_{\sigma =\sigma_i} 2\lambda_{\underline{\alpha} \underline{q}}(\sigma)\lambda_{\underline{\beta} \underline{q}}(\sigma)=
4\rho^{\#}_i v_{\underline{\alpha} \underline{A}i}^-v_{\underline{\beta} i}^{- \underline{A}}=
2\rho^{\#}_i v_{\underline{\alpha} \underline{q}i}^-v_{\underline{\beta} \underline{q}i}^-= k_i\!\!\!\!/_{\underline{\alpha}\underline{\beta}}$ and thus that Eq. (\ref{Pmu=lGl}) with
(\ref{Pmu=}) is satisfied. This is to say, the scattering equation (\ref{ScEqli=lj}) follows from Eqs. (\ref{Pmu=lGl}) with
(\ref{Pmu=})  and (\ref{sol=pScEq1}), \eqref{Wi=WiO}.


Notice that while the scattering equation (\ref{ScEq1}) is homogeneous, the polarized scattering equation (\ref{pScEq}) is not. As it is seen from its equivalent form \eqref{ScEqvi=vj1}, the scattering equation  provides a decomposition of $i-th$  helicity spinors $\bar{\lambda}_{\underline{\alpha} i}{}^{\underline{A}}$ (or complex spinor frame variables $\bar{v}{}_{\underline{\alpha} i}^{-\underline{A}}$), which provide the basis of the {\it complementary to} the space of common zero modes  of $k_i\!\!\!\!/$ and $U_i\!\!\!\!\!\!/\;$ in the space of solutions of the massless Dirac equation, on the set of the variables $\lambda_{\underline{\alpha} \underline{B}j}$ (or $v_{\underline{\alpha} \underline{B}j}^{\; -}$) providing the basis of the  spaces of common eigenfunctions of $k_j\!\!\!\!\!/\;$ and $U\!\!\!\! /{}_j\; $ with $j\not=i$.

\section{Supersymmetry generator and supersymmetric invariant amplitudes}
\label{secQ}

The supersymmetry generator can be realized as a differential operator in superspace with 11D Majorana spinor fermionic coordinate $\theta^{\underline{\alpha}}$ as well as in the real analytic superspace with
$16$ component Majorana spinor $$\theta^{-}_{\underline{q}}=\theta^{\underline{\alpha}}v_{\underline{\alpha}\underline{q}}^{\; -}$$ (see \cite{Bandos:2016tsm,Bandos:2017eof} and refs. therein). To this end the introduction of spinor frame
variables $v_{\underline{\alpha}\underline{q}}^{\; -}$ \eqref{harmV=11D} is necessary.
Furthermore, introducing also the internal frame variables \eqref{harm=SO9} parametrizing the coset $SO(9)/(SO(7)\times SO(2))$, one can construct a complex $8$--component fermionic coordinates $$\eta^-_{\underline{A}} = \theta^{-}_{\underline{q}} \bar{w}_{\underline{q}\underline{A}}$$ (see \cite{Bandos:2017zap}) and realize the supersymmetry generator as
\begin{eqnarray}\label{Qal=Qqv}
Q_{\underline{\alpha}} =
4\rho^{\#} v_{\underline{\alpha}}^{-\underline{A}}\eta^-_{\underline{A}}+
v_{\underline{\alpha}\underline{A}}^{\; -} \frac {\partial}{\partial  \eta^-_{\underline{A}}} =: v_{\underline{\alpha}\underline{q}}^{\; -}Q^+_{\underline{q}}\; .
\end{eqnarray}
We refer to \cite{Bandos:2017zap} and refs. therein for more details.

It is not difficult to check that \eqref{Qal=Qqv} obeys the superalgebra
\begin{eqnarray}\label{QQ=k}
{}\{ Q_{\underline{\alpha}}, Q_{\underline{\beta}}\} =
8\rho^{\#} \bar{v}{}_{(\underline{\alpha}}^{-\underline{A}}v_{\underline{\beta}) \underline{A}}^{\; -}= 4\rho^{\#} v_{\underline{\alpha}\underline{q}}^{\; -}v_{\underline{\beta} \underline{q}}^{\; -}&=&
2 \rho^{\#} u_{\underline{\mu}}^= \Gamma^{\underline{\mu}}_{\underline{\alpha}\underline{\beta}}\qquad \nonumber \\
&=&
2 k_{\underline{\mu}} \Gamma^{\underline{\mu}}_{\underline{\alpha}\underline{\beta}}\; .
\end{eqnarray}
This is the standard 11D supersymmetry algebra with the translation generator realized as 11D light-like momentum
(\ref{k=ru--}). Such a representation of the supersymmetry algebra was used in \cite{Geyer:2019ayz} so that our discussion here just clarifies the
meaning of the bosonic and fermionic variables used there and their relation with the ones used in \cite{Bandos:2016tsm,Bandos:2017zap,Bandos:2017eof}.

For the scattering problem the complete supersymmetry generator is given by the sum of 'partial' supersymmetry generators acting on the fermionic variables associated to different particles
\begin{eqnarray}\label{Qal=SQali}
Q_{\underline{\alpha}} =\sum\limits_i Q_{\underline{\alpha}i} =\sum\limits_i\left(
4\rho^{\#}_i \bar{v}_{\underline{\alpha}i}^{-\underline{A}}\eta^-_{\underline{A}i}+
v_{\underline{\alpha}\underline{A}i}^{\; -} \frac {\partial}{\partial  \eta^-_{\underline{A}i}}\right)\; .
\end{eqnarray}
It is nilpotent:  ${}\{ Q_{\underline{\alpha}}, Q_{\underline{\beta}}\} =0$ due to the momentum conservation.

Below we find convenient to use also the $SO(1,1)$ invariant fermionic variables
\begin{eqnarray}\label{eta=sqrtr-eta}
\eta_{\underline{A}i}:= \sqrt{\rho^{\#}_i} \eta^-_{\underline{A}i}\;
\end{eqnarray}
which is the supersymmetry partner of the complex helicity spinor
$\lambda_{\underline{\alpha} \underline{A}i}= \sqrt{\rho^{\#}_i}v_{\underline{\alpha} \underline{A}i}^{-}$ \eqref{lA:==11D},
\begin{eqnarray}\label{susy=etaAi}
\delta_\epsilon \eta_{\underline{A}i} = \epsilon^{\underline{\alpha}}
\lambda_{\underline{\alpha} \underline{A} i}
\; .
\end{eqnarray}
In terms of these and the helicity spinor variables the supersymmetry generator has the form of
\begin{eqnarray}\label{Qal=SQali=}
Q_{\underline{\alpha}} =\sum\limits_i Q_{\underline{\alpha}i} =\sum\limits_i\left(
4\bar{\lambda}{}_{\underline{\alpha}i}^{\;\;\underline{A}}\eta_{\underline{A}i}+
\lambda_{\underline{\alpha}\underline{A}i} \frac {\partial}{\partial  \eta_{\underline{A}i}}\right)\; .
\end{eqnarray}

The supersymmetric invariant $ e^{\,{\bb F}}$ found in  \cite{Geyer:2019ayz}
\begin{eqnarray}\label{QaleF=0}
Q_{\underline{\alpha}} e^{\,{\bb F}}=0
\end{eqnarray}
 is the exponent of
\begin{eqnarray}\label{F=}
{\bb F}&=& 2\sum\limits_i  \sum\limits_j \sqrt{\rho^{\#}_j\rho^{\#}_i} \frac {W^{\underline{A}}_{\underline{q}j}W^{\underline{B}}_{\underline{q}i}}{\sigma_j-\sigma_i}\eta^-_{\underline{A}j}\eta^-_{\underline{B}i}
\qquad \nonumber \\ &=& 2\sum\limits_i  \sum\limits_j  \frac {W^{\underline{A}}_{\underline{q}j}W^{\underline{B}}_{\underline{q}i}}{\sigma_j-\sigma_i}\eta_{\underline{A}j}\eta_{\underline{B}i}
\; .
\end{eqnarray}

The proof of the supersymmetric invariance of $ e^{\,{\bb F}}$ (\ref{QaleF=0}) passes through ({\it cf.} \cite{Geyer:2019ayz}, see  \eqref{Qal=SQali=})
\begin{eqnarray}\label{susyF=}
\sum\limits_i
\lambda_{\underline{\alpha}\underline{A}i}\frac {\partial}{\partial  \eta_{\underline{A}i}}{\bb F}= 4\sum\limits_i  \sum\limits_j  \frac {W^{\underline{A}}_{\underline{q}j} \lambda_{\underline{\alpha}\underline{A}j} }{\sigma_j-\sigma_i}W^{\underline{B}}_{\underline{q}i}\eta_{\underline{B}i}= \nonumber \\= 4\sum\limits_i  \sum\limits_j  \frac {W^{\underline{A}}_{\underline{q}j}(\sigma_i) \lambda_{\underline{\alpha}\underline{A}j} }{\sigma_j-\sigma_i}W^{\underline{B}}_{\underline{q}i}(\sigma_i)\eta_{\underline{B}i}= \nonumber \\ = -4 \sum\limits_i
 \lambda_{\underline{\alpha}\underline{q}}(\sigma_i)W^{\underline{B}}_{\underline{q}i}(\sigma_i) \eta_{\underline{B}i} = \nonumber \\
= - 4\sum\limits_i
\lambda_{\underline{\alpha}i}^{\underline{B}}\eta_{\underline{B}i}
\; .
\end{eqnarray}
Here the derivation of the first equality is straightforward,
to pass to the second line we have used  \eqref{WW=WsWs} (which is equivalent to
\eqref{Wi=WiO}), to arrive at the third  line we have used the expression  (\ref{sol=pScEq1}) for the meromorphic spinor function  and the fourth line is derived with the use of the polarized scattering equation (\ref{pScEq}).

The factor $e^{\,{\bb F}}$ determines the fermionic contribution to the superamplitude or S-matrix element.
In \cite{Geyer:2019ayz} it was proposed that this is given essentially by CHY expression \cite{Cachazo:2013hca} but with the factor $e^{\,{\bb F}}$  included into the integrand,
\begin{eqnarray}\label{S-mat=11D}
{\cal A}_n= \int \frac {1}{vol (SL(2,{\bb C}))} \, \prod\limits_{i=1}^{n}\; d\sigma_i\, \prod\limits_{i=1}^{n}{}^\prime \; \delta (k_i\cdot P(\sigma_i)) \det{}^\prime {\bb M}\; e^{\,{\bb F}}\; . \qquad
\end{eqnarray}
In this expression $k_i\cdot P(\sigma_i)= k_i{}^{\underline \mu} P_{\underline \mu}(\sigma_i)$,
\bea\label{prod'}
 \prod\limits_{i=1}^{n}{}^\prime \; \delta (k_i\cdot P(\sigma_i))  = \sigma_{jk} \sigma_{kl} \sigma_{lj} \prod\limits_{i=1, i\not=j,k,l}^{n} \; \delta (k_i\cdot P(\sigma_i))
\eea
is independent on choice of $j,k,l$, $\sigma_{ij}=\sigma_i-\sigma_j$,
${\bb M}$ is $2n\times 2n$ CHY matrix
\begin{eqnarray}\label{bbM=}
{\bb M}=\; \left(\begin{matrix} \frac {k_i\cdot k_j}{\sigma_{ij}} &\qquad & \frac {U_i\cdot k_j}{\sigma_{ij}} -
U_i\cdot P(\sigma_i)\delta_{ij} \qquad {}\cr & & \cr
{} \quad -\frac {U_j\cdot k_i}{\sigma_{ji}} +
U_j\cdot P(\sigma_j) \delta_{ij}\quad & \qquad & \frac {U_i\cdot U_j}{\sigma_{ij}} \end{matrix}\right)\; , \qquad
\end{eqnarray}
and
\begin{eqnarray}\label{det'=}
\det{}^\prime {\bb M}= \frac 4 {\sigma_{ij}^2} \det {\bb M}^{ij}_{ij}\; ,
\end{eqnarray}
where $ \det {\bb M}^{ij}_{ij}$ is the determinant of   $2(n-1)\times 2(n-1)$ matrix ${\bb M}^{ij}_{ij}$  obtained from \eqref{bbM=} by removing rows $i,j$ and columns $i,j$. Again, this latter is independent on choice of $i$ and $j$ \cite{Cachazo:2013hca}.

\section{Polarized scattering equation and spinor moving frame  formulation of  ambitwistor superstring in D=11}

The Green-Schwarz (or Brink---Schwarz) formulation of the ambitwistor superstring action is reached by considering the  Brink---Schwarz   superparticle Lagrangian, allowing in it all the fields to be dependent on two worldsheet coordinates, replacing  the proper time derivatives
$d/d\tau$ with holomorphic partial derivatives $\bar{\partial}$, and integrating it over the two dimensional worldsheet \cite{Bandos:2014lja}. In such a way we arrive at
\begin{eqnarray}\label{SaTw=s0}
S=\; \int_{{\cal W}^2}d^2\sigma \left( P_{\underline{\mu}} \left(\bar{\partial}X^{\underline{\mu}} -i \bar{\partial}\theta \Gamma^{\underline{\mu}} \theta \right)
-{e\over 2}  P^2\right) \; ,    \qquad
\end{eqnarray}
where $ P_{\underline{\mu}} (\sigma)$ is a vector density playing the role of the momentum conjugate to the bosonic coordinate function $X^{\underline{\mu}}(\sigma)$, $\theta^{\underline{\alpha}}(\sigma)$ are fermionic 32-component Majorana spinor coordinate function,
$\bar{\partial}\theta \Gamma^{\underline{\mu}} \theta=\bar{\partial}\theta^{\underline{\alpha}} \Gamma^{\underline{\mu}}_{\underline{\alpha}\underline{\beta}} \theta^{\underline{\beta}}$,
 and  $e (\sigma)$ is a Lagrange multiplier producing the constraint (\ref{P2=0}). Solving this constraint with the use of spinor frame fields related to $P_{\underline{\mu}} (\sigma)$ by (\ref{P=lGl}), or (\ref{Pmu=lGl}) and \eqref{l=r12v},  we arrive at the action of the spinor moving frame formulation of the 11D ambitwistor string \cite{Bandos:2014lja}.

This action can be written in an equivalent form \cite{Bandos:2014lja}
\begin{eqnarray}\label{SaTw=spf}
S&=&\; \int_{{\cal W}^2}d^2\sigma
\lambda_{\underline{\alpha}\underline{q}} (\sigma ) \lambda_{\underline{\beta}\underline{q}}(\sigma )
\left( \bar{\partial}X^{\underline{\alpha}\underline{\beta}}(\sigma ) -i \bar{\partial}\theta ^{(\underline{\alpha}}\,  \theta^{\underline{\beta})} (\sigma )\right)\;    \qquad \nonumber \\
&\equiv &\; \int_{{\cal W}^2}d^2\sigma
\rho^{\#}(\sigma )  v_{\underline{\alpha}\underline{q}}^{\; -}(\sigma )  v_{\underline{\beta}\underline{q}}^{\; -}(\sigma )
\left( \bar{\partial}X^{\underline{\alpha}\underline{\beta}}-i \bar{\partial}\theta ^{(\underline{\alpha}}\,  \theta^{\underline{\beta})} \right)\;    \qquad
\end{eqnarray}
with  an arbitrary symmetric spin tensor bosonic coordinate functions
\begin{eqnarray}
\label{Xalbe=}
X^{\underline{\alpha}\underline{\beta}}(\sigma)=X^{\underline{\beta}\underline{\alpha}}(\sigma)\equiv
\frac {1}{32} \tilde{\Gamma}_{\underline{\mu}}{}^{\underline{\alpha}\underline{\beta}} X^{\underline{\mu}}(\sigma)
 - {1\over 64} i Z^{\underline{\mu}\underline{\nu}} (\sigma)\tilde{\Gamma}_{\underline{\mu}\underline{\nu}}{}^{\underline{\alpha}\underline{\beta}} + {1\over {32\cdot\, 5!}}
 Z^{\underline{\mu}_1\ldots \underline{\mu}_5}(\sigma)
 \tilde{\Gamma}_{\underline{\mu}_1\ldots \underline{\mu}_5}{}^{^{\underline{\alpha}\underline{\beta}}} \; .  \quad
\end{eqnarray}
The properties of the spinor frame variables/helicity spinors concentrated in
\eqref{P=lGl}/\eqref{Pmu=lGl}
guarantee that the arbitrary variation of the
$  Z^{\underline{\mu}\underline{\nu}}(\sigma)$ and  $
 Z^{\underline{\mu}_1\ldots \underline{\mu}_5} (\sigma)$  do not change the action (see \cite{Bandos:2006nr} for the discussion in the context of massless superparticle model). This is the statement of gauge symmetry which can be fixed just by setting $Z^{\underline{\mu}\underline{\nu}}(\sigma )=0$ and  $
 Z^{\underline{\mu}_1\ldots \underline{\mu}_5}(\sigma)=0$ thus reducing (\ref{Xalbe=}) to
\begin{eqnarray}
\label{Xalbe=Xa}
X^{\underline{\alpha}\underline{\beta}}(\sigma)=
\frac {1}{32} \tilde{\Gamma}_{\underline{\mu}}^{\underline{\alpha}\underline{\beta}} X^{\underline{\mu}}(\sigma)\; .  \qquad
\end{eqnarray}
Just this gauge fixed form of the action \eqref{SaTw=spf}, with \eqref{Xalbe=Xa},  is related to \eqref{SaTw=s0} by the procedure described above. However, as we will see in a moment, it is sometimes convenient to treat the ambitwistor superstring as a dynamical system in the enlarged superspace $\Sigma^{(528|32)}$ with $528$ bosonic coordinates
$(X^{\underline{\mu}}, Z^{\underline{\mu}\underline{\nu}},
 Z^{\underline{\mu}_1\ldots \underline{\mu}_5})$ and $32$ fermionic coordinates $\theta{}^{\underline{\alpha}}$.

\subsection{Supertwistor formulation of the 11D ambitwistor superstring}

The action (\ref{SaTw=spf}) can be written as
\begin{eqnarray}\label{SaTw=sTw}
S=\int_{{\cal W}^2}d^2\sigma \left( \lambda_{\underline{\alpha}\underline{q}}\,  \bar{\partial} {\mu}^{\underline{\alpha}}_{\underline{q}}-
\bar{\partial}\lambda_{\underline{\alpha}\underline{q }}\;  {\mu}^{\underline{\alpha}}_{\underline{q}} - i \bar{\partial}\eta_{\underline{q}}\,\eta_{\underline{q}}\right)
\; ,
\end{eqnarray}
where
\begin{eqnarray}\label{lambda=}
 \lambda_{\underline{\alpha}\underline{q}}(\sigma)= \sqrt{\rho^{\#}(\sigma)}  v_{\underline{\alpha}\underline{p}}^{\;-}(\sigma){\cal S}_{\underline{p}\underline{q}} (\sigma)\; ,  \qquad
\end{eqnarray}
(see \eqref{l=r12v}) and
\begin{eqnarray}
\label{mu=}
{\mu}^{\underline{\alpha}}_{\underline{q}}(\sigma):= X^{\underline{\alpha}\underline{\beta}} (\sigma)\lambda_{\underline{\beta} \underline{q}}(\sigma) - {i\over
2} \theta^{\underline{\alpha}}(\sigma) \, \theta^{\underline{\beta}}(\sigma) \lambda_{\underline{\beta} \underline{q}}(\sigma)  \; , \qquad \\
\label{eta=}
\eta_{\underline{q}}(\sigma):= \theta^{\underline{\beta}}(\sigma) \lambda_{\underline{\beta} \underline{q}}(\sigma) \; .  \qquad
\end{eqnarray}
These are the 11D generalizations of the four dimensional Penrose incidence relations. They are imposed on the
set of $16$ constrained 11D supertwistors
$$
{\cal Z}{}_{{\Lambda} \underline{q}} = \left(\lambda_{\underline{\alpha} \underline{q}}\; ,\; {\mu}^{\underline{\alpha}}_{\underline{q}}\; , \;  \eta_{\underline{q}}\right)\;
$$
(see \cite{Bandos:2006nr} and refs. therein for more discussion on these).

Eqs. \eqref{eta=} and  (\ref{mu=}) with \eqref{Xalbe=}
describe the general solution of $120$ constraints
\begin{eqnarray}
\label{SO(16)=} & {\mathbb{J}}_{\underline{p}\underline{q}}
:=
 2 \lambda_{\underline{\alpha} [\underline{p}}{\mu}_{\underline{q}]}{}^{
\underline{\alpha}} + i {\eta}_{\underline{p}}{\eta}_{\underline{q}} =0 \,  \qquad
\end{eqnarray}
which can be identified with generator of $SO(16)$ gauge symmetry in the Hamiltonian formalism.

The rigid supersymmetry living invariant the action (\ref{SaTw=spf})
\begin{eqnarray}
\label{susy=X}  \delta_\epsilon X^{\underline{\alpha}\underline{\beta}} =i \theta^{(\underline{\alpha}}\epsilon^{\underline{\beta})}\, , \qquad \delta_\epsilon  \theta^{\underline{\alpha}}= \epsilon^{\underline{\alpha}}\, , \qquad
\end{eqnarray}
is realized on our constrained supertwistor by
\begin{eqnarray}
\label{susy=sTw}  \delta_\epsilon \lambda_{\underline{\alpha} \underline{q}}=0\, , \qquad \delta_\epsilon {\mu}_{\underline{q}}{}^{
\underline{\alpha}} =- i \epsilon^{
\underline{\alpha}} \eta_{\underline{q}}\, , \qquad \delta_\epsilon \eta_{\underline{q}}=  \epsilon^{
\underline{\alpha}} \lambda_{\alpha \underline{q}} \, . \qquad
\end{eqnarray}

Eq. (\ref{mu=})  with (\ref{Xalbe=Xa}) provides, together with \eqref{eta=}, the general solution of a bigger set of constraints including, besides \eqref{SO(16)=}, the set of $135$ constraints
\begin{eqnarray}
\label{moreII=} K_{\underline{p}\underline{q}}= K_{\underline{q}\underline{p}}:= \lambda_{\underline{\alpha} (\underline{p}}\;
{\mu}^{\underline{\alpha}}_{\underline{q})} - {1\over 16}\delta_{\underline{p}\underline{q}}\, \lambda_{\underline{\alpha}
\underline{p'}}\, {\mu}^{\underline{\alpha}}_{\underline{p'}}= 0 \; .
\end{eqnarray}
From the perspective of the system in enlarged superspace  $\Sigma^{(528|32)}$, these are  gauge fixing conditions for a  gauge symmetry which will be described below.

Thus, keeping in mind the generic form of spin--tensorial coordinate \eqref{Xalbe=}  in (\ref{mu=}) we can describe the 11D ambitwistor superstring by the action
\eqref{SaTw=sTw} with variables restricted by the constraints  (\ref{SO(16)=}) and (\ref{Pmu=lGl})
 \footnote{\label{footn-l} Here we mean that the light-like vector $P_\mu(\sigma)$ is defined by Eqs. (\ref{Pmu=lGl}) themselves. Alternatively one can state that $\lambda_{\underline{\alpha} \underline{q}}(\sigma)$ is restricted by the (reducible) set of the constraints
$$ \lambda_{\underline{p}}\Gamma^{\underline{\mu}\underline{\nu}}\lambda_{\underline{p}}=0\; , \qquad \lambda_{\underline{p}}\Gamma^{\underline{\mu}\underline{\nu}\underline{\rho}\underline{\sigma}\underline{\kappa}}\lambda_{\underline{p}}=0\; , \qquad
\lambda_{\underline{q}}\Gamma^{\underline{\mu}}\lambda_{\underline{p}}=\frac 1 {16} \delta_{\underline{q}\underline{p}} \, \lambda_{\underline{r}}\Gamma^{\underline{\mu}\underline{\nu}}\lambda_{\underline{r}}\; . \qquad
$$
}.

Furthermore,  we can introduce the constraint (\ref{SO(16)=}) with Lagrange multiplier into the action,
\begin{eqnarray}\label{SaTw=sTwL}
S=\int_{{\cal W}^2}d^2\sigma \left( \lambda_{\underline{\alpha}\underline{q}}\,  \bar{\partial} {\mu}^{\underline{\alpha}}_{\underline{q}}-
\bar{\partial}\lambda_{\underline{\alpha}\underline{q }}\;  {\mu}^{\underline{\alpha}}_{\underline{q}} - i \bar{\partial}\eta_{\underline{q}}\,\eta_{\underline{q}}\right)
+ \int_{{\cal W}^2}d^2\sigma \bar{{\cal A}}{}^{\underline{p}\underline{q}}\left( 2 \lambda_{\underline{\alpha} [\underline{p}}\mu^{
\underline{\alpha}}_{\underline{q}]} + i {\eta}_{[\underline{p}}{\eta}_{\underline{q}]}\right)
\;
\end{eqnarray}
and consider  the variables ${\mu}^{\underline{\alpha}}_{\underline{q}}$ as unconstrained.
It is important that the action \eqref{SaTw=sTwL} is invariant under $SO(16)$ gauge symmetry \eqref{l-lO} provided
 \begin{eqnarray}\label{mu-muO}
\mu^{\underline{\alpha}}_{ \underline{q}}(\sigma )\mapsto  \mu^{\underline{\alpha}}_{ \underline{p}}(\sigma ) {\cal O}_{\underline{p} \underline{q}}(\sigma )
\;  \qquad
 \qquad
\end{eqnarray}
and  the Lagrange multiplier
$ \bar{{\cal A}}{}^{\underline{p}\underline{q}}=\bar{{\cal A}}{}^{[\underline{p}\underline{q}]}$ is transformed as a gauge field under this symmetry,
\begin{eqnarray}
\label{vcalA}
\; \bar{{\cal A}}{}^{\underline{p}\underline{q}}\; \mapsto \; \left({\cal O}^{-1} \bar{\partial} {\cal O}+ {\cal O}^{-1} \bar{\cal A} {\cal O} \right)^{\underline{p}\underline{q}}\; .
\end{eqnarray}

The action (\ref{SaTw=sTw}) is also invariant under the following gauge symmetry transformations
\begin{eqnarray}
\label{vmu=gauge} \delta {\mu}^{\underline{\alpha}}_{\underline{q}}=   - {1\over 64} i \delta Z^{\underline{\nu}_1\underline{\nu}_2} (\sigma)\tilde{\Gamma}_{\underline{\nu}_1 \underline{\nu}_2}^{\underline{\alpha}\underline{\beta}}\lambda_{\underline{\beta} \underline{q}} + {1\over {32\cdot\, 5!}}
 \delta  Z^{\underline{\nu}_1\ldots \underline{\nu}_5}(\sigma)
 \tilde{\Gamma}_{\underline{\nu}_1\ldots \underline{\nu}_5}{}^{^{\underline{\alpha}\underline{\beta}}} \lambda_{\underline{\beta} \underline{q}}
\;   \quad
\end{eqnarray}
with arbitrary $\delta Z^{\underline{\mu}\underline{\nu}}(\sigma)$ and $ \delta  Z^{\underline{\nu}_1\ldots \underline{\nu}_5}(\sigma)$. This symmetry
allows for the gauge fixing conditions reducing the general solution (\ref{mu=}) of the constraints to
\begin{eqnarray}
\label{mu=11D}
{\mu}^{\underline{\alpha}}_{\underline{q}}:= \frac 1 {32} X^{\underline{\nu}} \tilde{\Gamma}_{\underline{\nu}}^{\underline{\alpha}\underline{\beta}} \lambda_{\underline{\beta} \underline{q}} - {i\over 2} \theta^{\underline{\alpha}} \, \theta^{\underline{\beta}} \lambda_{\underline{\beta} \underline{q}}\; . \qquad
\end{eqnarray}
This gauge is not preserved by supersymmetry transformations (\ref{susy=X}) along so that to reach the simple transformation of supertwistor (\ref{susy=sTw}) and to preserve the gauge (\ref{mu=11D}) one needs to supplement
(\ref{susy=X}) by the gauge transformations of the supertwistor (\ref{vmu=gauge}).

Of course,  the fields $\lambda_{\underline{\alpha}\underline{q}}(\sigma)$ are constrained by algebraic relation which follows from their expression in terms of spinor moving frame variables  (\ref{lambda=}) (these are actually collected in (\ref{Pmu=lGl}), see footnote \ref{footn-l}).
However, the fact that $ {\mu}^{\underline{\alpha}}_{\underline{q}}(\sigma)$ in the action (\ref{SaTw=sTwL}) can be treated as unconsrained will be very useful in our discussion below.

\subsection{11D ambitwistor superstring and polarized scattering equation}
\label{Ambi-Pol}


In the spinor frame formalism the SO(16) gauge invariant generalization of the vertex operator proposed in \cite{Geyer:2019ayz} reads
\begin{eqnarray}\label{V=int}
V= \int d^2\sigma_i \delta (k_i\cdot P(\sigma_i)) \frak{W} \exp \left(2i\mu^{\underline{\alpha}}_{\underline{q}} (\sigma_i)
\sqrt{\rho^{\#}_i} v_{\underline{\alpha}\underline{A}i}^{\; -} W_{\underline{q}i}^{\underline{A}}(\sigma_i) + \sqrt{\rho^{\#}(\sigma_i)} \theta^-_{\underline{q}}(\sigma_i)
\sqrt{\rho^{\#}_i} \eta^-_{\underline{A}i} W_{\underline{q}i}^{\underline{A}}(\sigma_i)
\right) \qquad \nonumber \\
=: \int d^2\sigma_i \delta (k_i\cdot P(\sigma_i)) \frak{W} \exp \left(2i\mu^{\underline{\alpha}}_{\underline{q}} (\sigma_i)
\lambda_{\underline{\alpha}\underline{A}i} W_{\underline{q}i}^{\underline{A}} (\sigma_i)+ 2\eta_{\underline{q}}(\sigma_i)
 \eta_{\underline{A}i} W_{\underline{q}i}^{\underline{A}}(\sigma_i)
\right)
\;   \qquad
\end{eqnarray}
where $\frak{W} $ denotes a possible additional worldsheet  operator depending on polarization data the explicit form of which will not be essential for our discussion (see \cite{Geyer:2019ayz} for further references describing its explicit form).
Besides this, the vertex operator \eqref{V=int} is expressed in terms of fermionic and  spinorial bosonic  functions describing the ambitwistor string,
 $\eta_{\underline{q}}(\sigma)$ and $\mu^{\underline{\alpha}}_{\underline{q}} (\sigma)$, $\lambda_{\underline{\alpha}\underline{q}} (\sigma)$
 (the latter entering  $\delta (k_i\cdot P(\sigma_i))$ where $P_\mu(\sigma)$ is assumed to be taken from
 \eqref{Pmu=lGl}), and the scattering data of $i$-th particle. These latter are described by $\lambda_{\underline{\alpha}Ai}$, which  defines
 $k_i$ through \eqref{kG=rvAvA} and polarization vector through \eqref{UluA=ldA}, fermionic $\eta_{Ai} = \sqrt{\rho^{\#}_i} \eta^-_{Ai}$ and bosonic matrix {\it function} $W_{\underline{q}i}^A (\sigma)$.

Despite of the  entrance of this latter into the set of scattering data, we consider it as a function of $\sigma_i$ to do not break explicitly the local $SO(16)$ symmetry characteristic for the ambitwistor superstring action \eqref{SaTw=sTwL}. On the other hand, the  entrance of $W_{\underline{q}i}^A (\sigma_i)$ into the set of scattering data suggests its identification with a constant matrices
 $W_{\underline{q}i}^A$ up to the universal ($i$-independent) local $SO(16)$  transformations, as described by
\eqref{Wi=WiO}. Furthermore, it also suggests the identification \eqref{Wi=wi} of the constant  matrices $W_{\underline{q}i}^A$ in \eqref{Wi=WiO} with  the internal frame matrix variable \eqref{harm=SO9} describing the polarization of the scattering particle through \eqref{Ug=bwbw}, so that  \eqref{Wi=WiO} becomes
 \begin{eqnarray}\label{W=wtS}
 W_{\underline{q}i}^A (\sigma) = w_{\underline{p}i}^A \tilde{{\cal O}}_{\underline{p}\underline{q}}(\sigma)  \;  ,  \qquad \tilde{{\cal O}}^T\tilde{{\cal O}}= {\bb I}_{16\times 16}\; .\qquad
\end{eqnarray}
As $\tilde{{\cal O}}_{\underline{p}\underline{q}} (\sigma) = \tilde{{\cal O}}_{\underline{q}\underline{p}}^{-1}(\sigma)$ is $SO(16)$ valued,  \eqref{W=wtS} would imply that  $W_{\underline{q}i}^A(\sigma) $ obeys, besides  the purity conditions,  also
\begin{eqnarray}\label{WbWs+cc=1}
&& W_{\underline{q}i}{}^{\underline{A}}(\sigma) \bar{W}_{\underline{p}\underline{A}i}(\sigma)+ \bar{W}_{\underline{q}\underline{A}i}(\sigma)W_{\underline{p}i}{}^{\underline{A}}(\sigma) =\delta_{\underline{q}\underline{p}}\; , \qquad \\
\label{bWWs=1}
&& \bar{W}_{\underline{q}\underline{B}i}(\sigma)W_{\underline{q}i}{}^{\underline{A}}(\sigma) =\delta_{\underline{B}}{}^{\underline{A}}\; , \qquad W_{\underline{q}i}{}^{\underline{A}}(\sigma) W_{\underline{q}i}{}^{\underline{B}}(\sigma) =0 \; , \qquad \bar{W}_{\underline{q}\underline{A}i} (\sigma) \bar{W}_{\underline{q}\underline{B}i} (\sigma)=0\;  \qquad
\end{eqnarray}
and thus describes an $SO(16)$ valued matrix field. Thus in the presence of vertex operators the $SO(16)$ symmetry is realized by St\"uckelberg mechanism.

The simplest calculations of the path integral with vertex operator insertions can be done by searching for a saddle point of the exponent of the action multiplied  by the exponential factors from vertex operators. This is to say, the main contribution to the path integral will come from the extrema of the action with the source terms coming from vertex operator. The essential for our purposes part of such an effective action reads
\begin{eqnarray}\label{SaTw+SV}
S+S_V=\int_{{\cal W}^2}d^2\sigma \left( \lambda_{\underline{\alpha}\underline{q}}\,  \bar{\partial} {\mu}^{\underline{\alpha}}_{\underline{q}}-
\bar{\partial}\lambda_{\underline{\alpha}\underline{q }}\;  {\mu}^{\underline{\alpha}}_{\underline{q}} -2 i \bar{\partial}\eta_{\underline{q}}\,\eta_{\underline{q}}\right)
+ \int_{{\cal W}^2}d^2\sigma \bar{{\cal A}}^{\underline{p}\underline{q}}\left( 2 \lambda_{\underline{\alpha} [\underline{p}}\mu^{
\underline{\alpha}}_{\underline{q}]} + i {\eta}_{[\underline{p}}{\eta}_{\underline{q}]}\right)+ \nonumber \\
+ \sum\limits_i \left(2\mu^{\underline{\alpha}}_{\underline{q}} (\sigma_i)
\sqrt{\rho^{\#}_i} v_{\underline{\alpha}\underline{A}i}^{\; -} W_{\underline{q}i}^{\underline{A}}(\sigma_i) -i \sqrt{\rho^{\#}(\sigma_i)} \theta^-_{\underline{q}}(\sigma_i)
\sqrt{\rho^{\#}_i} \eta^-_{\underline{A}i} W_{\underline{q}i}^{\underline{A}}(\sigma_i)\right)
\qquad
\nonumber \\
=\int_{{\cal W}^2}d^2\sigma \left( \lambda_{\underline{\alpha}\underline{q}}\,  \bar{\partial} {\mu}^{\underline{\alpha}}_{\underline{q}}-
\bar{\partial}\lambda_{\underline{\alpha}\underline{q }}\;  {\mu}^{\underline{\alpha}}_{\underline{q}} - i \bar{\partial}\eta_{\underline{q}}\,\eta_{\underline{q}}\right)
+ \int_{{\cal W}^2}d^2\sigma \bar{{\cal A}}^{[\underline{p}\underline{q}]}\left( 2 \lambda_{\underline{\alpha} [\underline{p}}\mu^{
\underline{\alpha}}_{\underline{q}]} + i {\eta}_{[\underline{p}}{\eta}_{\underline{q}]}\right)+ \nonumber \\
+ \sum\limits_i \int_{{\cal W}^2}d^2\sigma \delta(\sigma-\sigma_i) \left(2\mu^{\underline{\alpha}}_{\underline{q}} (\sigma)
\lambda_{\underline{\alpha}\underline{A}i} W_{\underline{q}i}^{\underline{A}}(\sigma)  -2i \eta_{\underline{q}}(\sigma)
\eta_{\underline{A}i} W_{\underline{q}i}^{\underline{A}}(\sigma)
\right)
\; . \qquad
\end{eqnarray}
It is invariant under the $SO(16)$ gauge symmetry and  contains $W_{\underline{q}i}^A(\sigma)$ which obeys \eqref{WqAWqB=0} and is assumed to be of the form \eqref{Wi=WiO}; moreover the fact that $W_{\underline{q}i}^A(\sigma)$ describes the scattering data suggests a more specific expresion
\eqref{W=wtS}. Clearly, no independent equation can be obtained by varying this St\"uckelberg field.

Equations of motion which follow from the variation of the action (\ref{SaTw+SV}) with respect to the unconstrained bosonic and fermionic fields,
${\mu}^{\underline{\alpha}}_{\underline{q}}(\sigma)$ and $\eta_{\underline{q}}(\sigma)$, have the form
\begin{eqnarray}\label{bDl=}
 \bar{D}\lambda_{\underline{\alpha}\underline{q}}(\sigma) &=& \sum\limits_i
 \delta(\sigma-\sigma_i)
\lambda_{\underline{\alpha}\underline{A}i} W_{\underline{q}i}^{\underline{A}} (\sigma_i)\; , \qquad
 \\
\label{bDeta=}
\bar{D}\eta_{\underline{q}}(\sigma) &=&  \sum\limits_i
 \delta(\sigma-\sigma_i)
\eta_{\underline{A}i} W_{\underline{q}i}^{\underline{A}} (\sigma_i)
\; , \qquad
\end{eqnarray}
where
\begin{eqnarray}\label{bDl:=}
\bar{D}\lambda_{\underline{\alpha}\underline{q}} = \bar{\partial}\lambda_{\underline{\alpha}\underline{q}}- \lambda_{\underline{\alpha}\underline{p}} \bar{{\cal A}}^{\underline{p}\underline{q}}
\; ,  \qquad \bar{D}\eta_{\underline{q}} = \bar{\partial}\eta_{\underline{q}}- \eta_{\underline{p}} \bar{{\cal A}}^{\underline{p}\underline{q}}
  \qquad
\end{eqnarray}
are $SO(16)$ covariant derivatives constructed with the use of Lagrange multiplier  $ \bar{{\cal A}}^{\underline{p}\underline{q}}$ as  SO(16) gauge field. Furthermore, this is a one component gauge field associated to the derivative in one (anti-holomorphic) complex direction
and, as such, it can always be gauged away.
In the gauge  \begin{eqnarray}\label{cA=0g} \bar{{\cal A}}^{\underline{p}\underline{q}}=0   \qquad
\end{eqnarray} the equations (\ref{bDl=}) and (\ref{bDeta=}) simplify to
\begin{eqnarray}\label{bdl=}
 \bar{\partial}\lambda_{\underline{\alpha}\underline{q}}(\sigma) &=& \sum\limits_i
 \delta(\sigma-\sigma_i)
\lambda_{\underline{\alpha}\underline{A}i} W_{\underline{q}i}^{\underline{A}}\; , \qquad
 \\
\label{bdeta=}
 \bar{\partial} \eta_{\underline{q}}(\sigma) &=&   \sum\limits_i
 \delta(\sigma-\sigma_i)
\eta_{\underline{A}i} W_{\underline{q}i}^{\underline{A}}
\; , \qquad
\end{eqnarray}
where we have assumed that
\begin{eqnarray}\label{Wi=WsiOsi}
{W}_{\underline{q} i}^{\; \underline{A}}\,= \tilde{\cal O}_{\underline{q} \underline{p}}(\sigma_i )  {W}_{\underline{p} i}^{\; \underline{A}} (\sigma_i )  \; \qquad
\end{eqnarray}
is independent on $\sigma_i$. This assumption is equivalent to  \eqref{Wi=WiO}; we also keep in mind the identification \eqref{Wi=wi} of this constant matrix with the internal harmonics providing the square root of (the conjugate to) the polarization vector
\eqref{Ug=bwbw}, $ \bar{U}\!\!\!\!/{}_{\underline{q}\underline{p}i}:= \bar{U}^{\underline{I}}_i \gamma^{\underline{I}}_{\underline{q}\underline{p}}= 2 {w}_{\underline{q}i}^{\; \underline{A} }{w}_{\underline{p}i}^{\; \underline{A} }$.

In \eqref{Wi=WsiOsi} $\tilde{\cal O}_{\underline{p}\underline{q}}(\sigma) $ is $SO(16)$ valued matrix field trivializing the connection given by the
Lagrange multiplier in the action \eqref{SaTw=sTwL},
\begin{eqnarray}
\label{calA=OdO}
\; \bar{{\cal A}}{}^{\underline{p}\underline{q}}= \left(\tilde{\cal O}^{-1} \bar{\partial} \tilde{\cal O} \right){}^{\underline{p}\underline{q}}\; .
\end{eqnarray}
Clearly, this matrix field corresponds to the gauge transformation which is used to fix the gauge \eqref{cA=0g}.

The solutions of the equations \eqref{bdl=} and \eqref{bdeta=} are given by
\begin{eqnarray}\label{sol=ScEqBg}
\lambda_{\underline{\alpha} \underline{q}}(\sigma )= \sum_{i=1}^n \frac { \lambda_{\underline{\alpha} \underline{A}i}{W}_{\underline{q} i}^{\; \underline{A}}}{\sigma-\sigma_i}\; , \\
\label{sol=ScEqFg}
\eta_{\underline{q}}(\sigma )=  \sum_{i=1}^n\frac { \eta_{\underline{A}i}{W}_{\underline{q} i}^{\; \underline{A}} }{\sigma-\sigma_i}\;.
\end{eqnarray}
These equations, which essentially coincide with ones presented in \cite{Geyer:2019ayz},  are invariant under the rigid $SO(16)$ symmetry only.

The solution of the gauge covariant equations \eqref{bDl=} and \eqref{bDeta=} can be obtained by performing the local $SO(16)$ transformations of \eqref{sol=ScEqBg} and \eqref{sol=ScEqFg} with matrices  $\tilde{\cal O}_{\underline{p}\underline{q}}(\sigma)$  related to the antiholomorphic component of the gauge field by \eqref{calA=OdO}.
This solution reads
\begin{eqnarray}\label{sol=ScEqB}
\lambda_{\underline{\alpha} \underline{q}}(\sigma )= \sum_{i=1}^n  \frac { \lambda_{\underline{\alpha}Ai}{W}_{\underline{q} i}^{\; A} (\sigma)}{\sigma-\sigma_i}\; , \\
\label{sol=ScEqF}
\eta_{\underline{q}}(\sigma )=  \sum_{i=1}^n  \frac { \eta_{Ai}{W}_{\underline{q} i}^{\; A} (\sigma)}{\sigma-\sigma_i}\; ,
\end{eqnarray}
where (see  Eq. \eqref{Wi=WiO})
\begin{eqnarray}\label{Ws=WiOs}{W}_{\underline{q} i}{}^{\underline{A}} (\sigma)= {W}_{\underline{p} i}^{\; \underline{A}}\; \tilde{\cal O}_{\underline{p} \underline{q}}(\sigma )\; .  \end{eqnarray}

If accepting the identification \eqref{Wi=wi}, which implies \eqref{W=wtS},  substituting that into \eqref{sol=ScEqB} and \eqref{sol=ScEqF} and using \eqref{Ug=bwbw} we obtain the expression for the bosonic spinor  and fermionic functions
in terms of real helicity spinors and polarization vectors
\begin{eqnarray}\label{sol=ScEqB-U}
\lambda_{\underline{\alpha} \underline{q}}(\sigma )= \sum_{i=1}^n  \frac { \lambda_{\underline{\alpha}\underline{p}i}
({U}\!\!\!\!/{}_i\bar{U}\!\!\!\!/{}_i){}_{\underline{p}\underline{q}^\prime }}{4(\sigma-\sigma_i)}\tilde{\cal O}_{\underline{q}^\prime \underline{q}}(\sigma )\; , \\
\label{sol=ScEqF-U}
\eta_{\underline{q}}(\sigma )=  \sum_{i=1}^n  \frac { \eta_{\underline{\alpha}\underline{p}i}
({U}\!\!\!\!/{}_i\bar{U}\!\!\!\!/{}_i{})_{\underline{p}\underline{q}' }}{4(\sigma-\sigma_i)}\tilde{\cal O}_{\underline{q}' \underline{q}}(\sigma )\;
\end{eqnarray}
with the same $\tilde{\cal O}_{\underline{p} \underline{q}}(\sigma )$ as in
\eqref{W=wtS}.

The polarized scattering equation (\ref{ScEqli=lj}) should be imposed on the scattering data thus producing its equivalent form (\ref{pScEq}) when the solution (\ref{sol=ScEqB}) of the ambitwistor string equations of motion is taken into account.
Other way arround is to say that, as we have already discussed, the polarized scattering equation (\ref{ScEqli=lj}) can be obtained as consistency conditions of the constraints \eqref{Pmu=lGl} with (\ref{sol=ScEqB}) and (\ref{Pmu=}).

The first of two equations which we have obtained from the ambitwistor superstring action, Eq. \eqref{sol=ScEqB}, coincides with the $SO(16)$ covariant ansatz (\ref{sol=pScEq1}) for the solution of the polarized  scattering equation \eqref{pScEq} which generalizes the ansatz of \cite{Geyer:2019ayz}.
The second equation, \eqref{sol=ScEqF}, provides the fermionic superpartner of \eqref{sol=ScEqB}.

Indeed, taking into account \eqref{susy=etaAi} and \eqref{susy=sTw}, one can check that the supersymmetry variation of  Eq. \eqref{sol=ScEqF} is proportional to  Eq. \eqref{sol=ScEqB},
\be\label{susyPSEq}
\delta_\epsilon \left(\eta_{\underline{q}}(\sigma )- \sum_{i=1}^n \sqrt{\rho^\#_i}\, \frac { \eta_{\underline{A}i}^{\; -}{W}_{\underline{q} i}^{\; \underline{A}} (\sigma)}{\sigma-\sigma_i}\right) = \epsilon{}^{\,\underline{\alpha} }\left(\lambda_{\underline{\alpha} \underline{q}}(\sigma )- \sum_{i=1}^n  \frac { \lambda_{\underline{\alpha}\underline{A}i}{W}_{\underline{q} i}^{\; \underline{A}} (\sigma)}{\sigma-\sigma_i}\right)\; ,
\ee
and hence vanishes due to this equation. As a result the system of equations  \eqref{sol=ScEqB} and \eqref{sol=ScEqF} is supersymmetric invariant.

\section{Fermionic superpartner of the polarized scattering equation}
\label{sec=fpScEq}

Thus, interestingly enough, in the ambitwistor superstring approach the meromorphic spinor function \eqref{sol=ScEqBg} appears accompanied by its fermionic superpartner \eqref{sol=ScEqFg}. This makes   tempting to search also for the fermionic superpartner of the polarized scattering equation. Just formally, the structure of the bosonic polarized scattering equation considered together with the knowledge on the origin of the complex fermionic variables
$\eta_A= \eta_q \bar{w}_{qA}=\theta^\alpha \lambda_{\alpha q} \bar{w}_{qA}$ (see \cite{Bandos:2017zap}) suggests to propose on this r\^ole
\begin{eqnarray}\label{spScEq=}
\eta_{\underline{q}}(\sigma_i ) {W}_{\underline{q}i}^{\underline{A}}(\sigma_i ) =   \eta_{i}^{\; \underline{A}}\; ,
\end{eqnarray}
where $\eta_{i}^{\; \underline{A}}= \eta_{\underline{q}i} {w}_{\underline{q}i}^A=\theta_i^{\underline{\alpha}} \lambda_{\underline{\alpha}\underline{q}i} {w}_{\underline{q}i}^{\underline{A}}$.
Indeed, it is easy to check that $$\delta_\epsilon (\eta_{\underline{q}}(\sigma_i ) {W}_{\underline{q}i}^{\underline{A}}(\sigma_i ) -   \eta_{i}^{\; \underline{A}})=\epsilon^{\underline{\alpha} }\left(
\lambda_{\underline{\alpha} \underline{q}}(\sigma_i ) {W}_{\underline{q}i}^{\underline{A}}(\sigma_i ) -\lambda_{\underline{\alpha}i}^{\; \underline{A}} \right)$$ so that Eq. \eqref{spScEq=} is supersymmetric invariant if the polarized scattering equation  \eqref{pScEq} holds.

However, literally \eqref{spScEq=}  does not feet in the polarized scattering equation formalism as far as in it the fermionic variables of $i$-th particle are described by complex $\eta_{i\underline{A}}$ while its complex conjugate $\eta_{i}^{\; \underline{A}}$ should be realized as differential operator
(see the expression for supersymmetry generators in sec. \ref{secQ} and \cite{Bandos:2017zap} for more details). Then, schematically, the proposed fermionic superpartner of the polarized scattering equation should read    \begin{eqnarray}\label{spScEq=f=}
\eta_{\underline{q}}(\sigma_i ) {W}_{\underline{q}i}^{\underline{A}}(\sigma_i ) =\frac 1 4  \frac \partial {\partial \eta_{i\underline{A}}}\;
\end{eqnarray}
and might be realizable as an equation imposed on the superamplitude (the value of the coefficient in the {\it r.h.s.} will become clear in no time).

 This is indeed the case. Taking into account the expression for the fermionic meromorphic function \eqref{sol=ScEqF}, we can easily find that ${\bb F}$ from \eqref{F=} satisfies  $\frac \partial {\partial \eta_{i\underline{A}}} {\bb F} = 4\eta_{\underline{q}}(\sigma_i ) {W}_{\underline{q}i}^{\underline{A}}(\sigma_i )$ so that the supersymmetric invariant $\exp {\bb F}$ obeys
 \begin{eqnarray}\label{fpScEeF=0}
\left( \frac \partial {\partial \eta_{i\underline{A}}}-4\eta_{\underline{q}}(\sigma_i ) {W}_{\underline{q}i}^{\underline{A}}(\sigma_i ) \right) e^{{\bb F}} =0 \; .
\end{eqnarray}

We can  use \eqref{sol=ScEqF}
to write \eqref{fpScEeF=0} in an equivalent form
$$\left( \frac \partial {\partial \eta_{i\underline{A}}}-4\sum\limits_{j=1, j\not=i }^n \frac { \eta_{j\underline{B}}{W}_{\underline{q}j}^{\underline{B}} {W}_{\underline{q}i}^{\underline{A}}}{\sigma_i-\sigma_j}\right) e^{{\bb F}} =0\; . $$
Such a form is convenient to search for  the equation obeyed by the   tree amplitude of 11D supergravity: it is not difficult to check that \eqref{S-mat=11D} satisfies
\begin{eqnarray}\label{feq-S-mat=11D}
\left( \frac \partial {\partial \eta_{i\underline{A}}}-4\sum\limits_{j=1, j\not=i }^n \frac { \eta_{j\underline{B}}{W}_{\underline{q}j}^{\underline{B}} {W}_{\underline{q}i}^{\underline{A}}}{\sigma_i-\sigma_j}\right){\cal A}_n^{11D}= 0\; . \qquad
\end{eqnarray}

Thus we have found the superpartner of the polarized scattering equation
\eqref{ScEqli=lj} which happens to be an equation imposed on the supergravity amplitude, Eq. \eqref{feq-S-mat=11D}.

\section{Spinor helicity formalism, polarized scattering equations and ambitwistor superstring in D=10}

In this section we will describe the spinor frame approach to 10D polarized scattering equation and its ambitwistor superstring origin. The similarity with 11D case will allow us to be brief; we will especially notice the stages where the differences between 10D and 11D cases appear.

\subsection{Spinor frame approach to 10D spinor helicity formalism I. Real helicity spinors }

Ten dimensional Lorentz harmonics $v_{{\alpha}\dot{q}}^{\; +}, v_{{\alpha} {q}}^{\; -}$ were introduced in \cite{Galperin:1991gk,Delduc:1991ir} and used to construct the spinor moving frame formulation of 10D Green-Schwarz superstring in \cite{BZ-str}   and superembedding approach in \cite{Bandos:1995zw} (see  \cite{Sorokin:1999jx} for a nice review).
They are rectangular $16\times 8$ blocks of the $16\times 16$ spinor frame matrix
\begin{eqnarray}\label{harmV=10D}
V_{{\alpha}}^{({\beta})}= \left(\begin{matrix} v_{{\alpha}\dot{q}}^{\; +} , & v_{{\alpha} {q}}^{\; -}
  \end{matrix}\right) \in Spin(1,9)\;  \qquad
\end{eqnarray}
carrying different $SO(1,1)$ weights ($\pm$) and the indices of different (c- and s-spinor) representations of the $SO(8)$ subgroup, $\dot{q}=1,...,8$ and $q=1,...,8$.
They also carry the Majorana-Weyl spinor index $\alpha=1,...,16$ of the 10D Lorentz group.

As there is no charge conjugation matrix in 10D Majorana-Weyl spinor representation, there is no Lorentz covariant manner to rise and to lower $Spin(1,9)$ indices. The position of spinor index of a field carries  the physical information on its chirality. In our case this fact implies that it is impossible to construct (in a Lorentz covariant manner)  the elements of the inverse of the spinor moving frame matrix
\begin{eqnarray}\label{harmV-1=10D}
V_{(\beta)}^{\;\;\; \alpha}= \left(\begin{matrix}  v^{+\alpha}_{{q}}
 \cr  v^{-\alpha}_{{\dot{q}}} \end{matrix} \right) \in Spin(1,9)
 \;  \qquad
\end{eqnarray}
from the above moving frame variables \eqref{harmV=10D} ({\it cf.} 11D case in \eqref{V-1=CV-A}). Hence we have to introduce them as independent variables and subject these to the constraints\footnote{This is similar to introduction of the inverse tetrade in general relativity.}
\begin{eqnarray}\label{v-qv+p=10D}
&
v^{+\alpha}_{{q}} v_{\alpha {p}}^{\; -} = \delta_{qp}
 \; ,  \qquad & v^{+\alpha}_{{q}}  v_{\alpha \dot{p}}^{\; +}=0\;  , \qquad
 \nonumber  \\
 &
 v^{-\alpha}_{\dot{q}}   v_{\alpha q}^{\; -}=0\;  , \qquad & v^{-\alpha}_{\dot{q}}   v_{\alpha \dot{p}}^{\; +}=\delta_{\dot{q}\dot{p}}\;   \qquad
\end{eqnarray}
which are tantamount to  $V_{(\beta)}{}^\gamma V_{\gamma}^{(\alpha)}=\delta _{(\beta)}{}^{(\alpha)}$.
The equation  $V_{\alpha}^{(\beta)}V_{(\beta)}{}^\gamma :=  v_{\alpha \dot{q}}^{\; -} v^{+\gamma}_{\dot{q}}
+ v^{-\alpha}_q v^{-\gamma}_q
=\delta_{\alpha}^\gamma$ is also valid as a consequence of \eqref{v-qv+p=10D}.

Both the spinor frame and inverse spinor frame variables (spinor harmonics) can be considered as square roots of the same vector frame variables (vector harmonics) defined as elements of the $SO(1,9)$ valued matrix
\begin{eqnarray}\label{Uab=in10D}
 u_{{\mu}}^{({a})} = \left( {1\over 2}\left( u_{{\mu}}^{=}+u_{{\mu}}^{\#}
 \right), \; u_{{\mu}}^{{I}} \, , {1\over 2}\left( u_{{\mu}}^{\#}-u_{{\mu}}^{=}
 \right)\right)\; \in \; SO^\uparrow (1,9)\,    \qquad \nonumber \\ \Leftrightarrow \qquad \begin{cases}  u_{{\mu}}^{=} u^{\mu=}=0\; , \quad u_{{\mu}}^{=} u^{\mu\#}=2\; , \quad  u_{{\mu}}^{\#} u^{\mu\#}=0\; , \;\cr  u_{{\mu}}^{=} u^{\mu I}=0\; , \quad u_{{\mu}}^{\#} u^{\mu I}=0\; , \quad  u_{{\mu}}^{I} u^{\mu J}=-\delta^{IJ}\; . \;  \end{cases}
\end{eqnarray}
In particular both $v_{\alpha q}{}^-$ and $v^{-\alpha}_{\dot q}$ can be considered as square roots of the same light-like vector
$u_{{\mu}}^{=}$ of the associated vector frame in the sense of
\begin{eqnarray}\label{10Du==v-v-}
 & u_\mu^= \sigma^\mu_{\alpha\beta}= 2v_{\alpha q}{}^- v_{\beta q}{}^-  \; , & \qquad
 v^-_{{q}} \tilde{\sigma}_{\mu}v^-_{{p}}= u_\mu^= \delta_{{q}{p}}, \qquad \\ \label{10Du==v-v-=} & u_\mu^= \tilde{\sigma}{}^{\mu\, \alpha\beta}= 2v^{-\alpha}_{\dot q}   v^{-\beta}_{\dot q} \; , & \qquad  v^-_{\dot{q}} {\sigma}_{\mu}v^-_{\dot{p}}= u_\mu^= \delta_{\dot{q}\dot{p}}
 \; . \qquad
\end{eqnarray}
Here $ \sigma^a_{\alpha\beta}$ and $\tilde{\sigma}{}^{a\, \alpha\beta}$ are 10D generalized Pauli matrices which obey
\begin{eqnarray}\label{satsb=}
\sigma_{\mu}\tilde{\sigma}_{\nu}+ \sigma_{\nu} \tilde{\sigma}_{\mu}= \eta_{\mu\nu} {\bb I}_{16\times 16}\; . \qquad
\end{eqnarray}
Relations \eqref{10Du==v-v-} and \eqref{10Du==v-v-=} also contain all the essential constraints obeyed by the spinor frame variables with negative $SO(1,1)$ weight,   $v_{\alpha q}{}^-$ and $v^{-\alpha}_{\dot q}$.
More details on 10D spinor frame variables suitable for the description of massless superparticle can be found e.g. in \cite{Bandos:2017zap,Bandos:2017eof} and in Appendix C.

Adapting the vector frame to the light-like momentum $k_{\mu }$ by orienting in its direction one of the light-like vectors of the frame,  say $u_\mu^=$,
\be\label{ka=u--a10D}
k_{\mu i}= \rho^\#_i u_{\mu i}^=\; ,
\ee
we can then relate this $k_{\mu }$ to left- and to right-handed helicity spinors
\begin{eqnarray}\label{l=sqrv10D}
\lambda_{{\alpha}{q} i}=  \sqrt{\rho^\#_i} v^-_{{\alpha} {q} i}\; ,   \qquad \lambda^{\; \alpha}_{\dot{q} i}=  \sqrt{\rho^\#_i} v^{-\alpha}_ {\dot{q} i}\;   \qquad
\end{eqnarray}
by
\begin{eqnarray}\label{kI=lsl}
 & k_\mu \sigma^\mu_{\alpha\beta}= 2\lambda_{\alpha q}\lambda_{\beta q}  \; , & \qquad
 \lambda_{{q}} \tilde{\sigma}_{\mu}\lambda_{{p}}= k_\mu \delta_{{q}{p}}, \qquad \\ \label{kI=ltsl} & k_\mu \tilde{\sigma}{}^{\mu\, \alpha\beta}= 2\lambda^{\;\alpha}_{\dot q}   \lambda^{\;\beta}_{\dot q} \; , & \qquad \lambda_{\dot{q}} {\sigma}_{\mu}\lambda_{\dot{p}}= k_\mu \delta_{\dot{q}\dot{p}}
 \; . \qquad
\end{eqnarray}
We can reverse the line of arguing and define the helicity spinors by Eqs. \eqref{kI=lsl} and \eqref{kI=ltsl}. Then for instance,
\begin{eqnarray}\label{ll=0}
\lambda^{\;\alpha}_{\dot p i}    \lambda_{\alpha q i}=0 \;  \qquad
\end{eqnarray}
follows from the light-likeness of the momentum, $k_{\mu i}k^\mu_i=0$ and the general solution of  \eqref{kI=lsl} and \eqref{kI=ltsl} can be written in the form of \eqref{l=sqrv10D}.

\subsection{Spinor frame approach to 10D spinor helicity formalism II. Internal frame and complex helicity spinors }

The state of a scattered  vector particle can be characterized by the momentum and a  complex polarization vector
 $U_{\mu i}$ which obeys
\begin{eqnarray}\label{10DUk=0}
 k_{\mu i} U_i^{\mu}=0\; , \qquad  U_{\mu i} U_i^{\mu}=0\; .  \qquad
\end{eqnarray}
As in 11D case we can decompose this on the spacelike vectors of the moving frame (\ref{Uab=in10D})
\begin{eqnarray}\label{10DU=UIuI}
U_{\mu i}= u^{I}_{\mu i}U^{I}_{i}\; , \qquad U^{I}_{i}U^{I}_{i}=0\; .
\;  \qquad
\end{eqnarray}
The coefficient $U^{I}_{i}$ is a complex null $SO(8)$ vector the presence of which breaks little group
of the D=10 massless particle $SO(8)$ down to tiny group $SU(4)$
(more precisely, to $SO(2)\otimes SO(6)=U(1)\otimes SU(4)$; see  \cite{CaronHuot:2010rj} and \cite{Bandos:2017eof} for more discussion). This null vector can be considered as a part of internal $SO(8)$ vector frame  and factorized as follows
\begin{eqnarray}
 \label{Ug8=bww}
& U\!\!\!\!/{}_{q \dot{p}}:=  \gamma^I_{q \dot{p}} U_I = 2  \bar{w}_{{q}A} w_{\dot{p}}^A  \; , & \qquad \bar{w}_{pA}\gamma^{I}_{p\dot q}w_{\dot q}{}^B= U^I \delta_A{}^B \;  \qquad
\end{eqnarray}
in terms of the elements of associated s-spinor and c-spinor frames \cite{Bandos:2017zap}
\begin{eqnarray}\label{bwwSO8}
\left(\begin{matrix} \bar{w}_{qA} \; , \; w_q^{A} \end{matrix}\right)\; \in \; SO(8) \; , \qquad
\left(\begin{matrix}  \bar{w}_{\dot{p}B} \; ,\;  w_{\dot{p}}^{B} \end{matrix}\right)\; \in \; SO(8)  \; . \qquad
\end{eqnarray}

These can be used also to form  the complex helicity spinors
\begin{eqnarray}\label{lA:==}
 \lambda_{\alpha A}:= \lambda_{\alpha q} \bar{w}_{qA} = \sqrt{\rho^\#} v_{\alpha q}^{-} \bar{w}_{qA}\; , \qquad \bar{\lambda}{}_{\alpha}^{\;A}:= \lambda_{\alpha {p}} {w}_{{p}}^{\; A}= \sqrt{\rho^\#} v_{\alpha {p}}^{-} {w}_{{p}}^{\; A} \; , \qquad  \\
\label{lA:==-1}   \lambda_{ A}^{\; \alpha}:= \lambda_{\dot{q}}^{\; \alpha} \bar{w}_{\dot{q}A}= \sqrt{\rho^\#} v_{\dot{q}}^{-\alpha} \bar{w}_{\dot{q}A} , \qquad \bar{\lambda}{}^{\; A\alpha }:= \lambda_{\dot{q}}^{\; \alpha} {w}_{\dot{q}}^{\; A} = \sqrt{\rho^\#} v_{\dot{q}}^{-\alpha} {w}_{\dot{q}}^{\; A}   \qquad
\end{eqnarray}
which encode more explicitly the information about polarization of massless 10D particles.

These complex spinors solve the left- and right-chiral versions of the Dirac-Weyl equation
\bea\label{tklA=010D}
\tilde{k}\!\!\!/{}^{\alpha\beta}_i \lambda_{\beta Ai}=0\; , \qquad
\tilde{k}\!\!\!/{}^{\alpha\beta}_i  \bar{\lambda}_{\beta i}^{\; A}=0\; , \qquad  \\ \label{klA=010D}
{k}\!\!\!/{}_{\alpha\beta\; i} \lambda_{Ai}^{\; \beta }=0\; , \qquad
{k}\!\!\!/{}_{\alpha\beta\; i} \bar{\lambda}_{ i}^{\; \beta A}=0\; , \qquad
\eea
while only a half of them are in the kernel of the matrices constructed from the polarization vector
\begin{eqnarray}\label{Us=vv} \tilde{U}\!\!\!\!/{}^{\alpha\beta}:= U_{\mu} \tilde{\sigma}{}^{\mu\alpha\beta}=
- 4
{v}^{+(\alpha}_{A}v^{-\beta) A}
\;  , \qquad U\!\!\!\!/{}_{\alpha\beta}:= U_{\mu} \sigma^{\mu}_{\alpha\beta}=
4
{v}^{\;-}_{(\alpha|A}v_{|\beta)}^{+A}
\;  . \qquad
\end{eqnarray}
Namely,
\bea\label{tUlA=10D}
\tilde{U}\!\!\!\!/{}^{\alpha\beta}_i \lambda_{\beta Ai}=0\; , \qquad
\tilde{U}\!\!\!\!/{}^{\alpha\beta}_i  \bar{\lambda}_{\beta i}^{\; A}=-2 \bar{\lambda}_{ i}^{\; \alpha A}\; , \qquad  \\ \label{UlA=10D}
{U}\!\!\!\!/{}_{\alpha\beta\; i} \lambda_{Ai}^{\; \beta }=- 2{\lambda}_{  \alpha A i}\; , \qquad
{U}\!\!\!\!/{}_{\alpha\beta\; i} \bar{\lambda}_{ i}^{\; \beta A}=0\; , \qquad
\eea

Thus ${\lambda}_{  \alpha A i}$  provide a basis of common zero modes of
$\tilde{k}\!\!\!/{}^{\alpha\beta}_i$ and $\tilde{U}\!\!\!\!/{}^{\alpha\beta}_i$ matrices while $ \bar{\lambda}_{\beta i}^{\; A}$ is the basis of complementary to the above space in the space of solutions of left-chiral Dirac equation. In the case of ${k}\!\!\!/{}_{\alpha\beta\; i} $ and ${U}\!\!\!\!/{}_{\alpha\beta\; i}$ matrices the same roles are played by $ \bar{\lambda}_{ i}^{\; \alpha A}$ and  $\lambda_{Ai}^{\; \alpha }$, respectively.

From \eqref{kI=lsl} and \eqref{kI=ltsl} one finds the following factorization of the Dirac-Weyl matrices of different chirality in terms of complex helicity spinors
\begin{eqnarray}\label{kI=lsl=10D}
 &  {k}\!\!\!/{}_{\alpha\beta\; i} := k_\mu \sigma^\mu_{\alpha\beta}= 4\lambda_{(\alpha| A}\bar{\lambda}_{|\beta )}^{\; A}  \; , & \qquad \tilde{k}\!\!\!/{}^{\alpha\beta}_i:= k_\mu \tilde{\sigma}{}^{\mu\, \alpha\beta}= 4\lambda^{(\alpha}_{A}   \bar{\lambda}^{\beta)A}
 \; . \qquad
\end{eqnarray}
The other constraints on the complex spinors following from \eqref{kI=lsl} and \eqref{kI=ltsl} read
\begin{eqnarray}\label{ll=ks}
 \lambda_{A} \tilde{\sigma}_{\mu}\bar{\lambda}^{B}= k_\mu \delta_{A}{}^{B}, \qquad \lambda_{A} \tilde{\sigma}_{\mu}{\lambda}_{B}= 0 , \qquad \bar{\lambda}^A \tilde{\sigma}_{\mu}\bar{\lambda}^{B}= 0, \qquad  \\ \label{ll=kts}  \lambda_{A} {\sigma}_{\mu}\bar{\lambda}^{B}= k_\mu \delta_{A}{}^{B}
 \; , \qquad \lambda_{A}{\sigma}_{\mu}{\lambda}_{B}= 0 , \qquad \bar{\lambda}^A {\sigma}_{\mu}\bar{\lambda}^{B}= 0. \qquad
\end{eqnarray}
These indicate, in particular, that  both the left chiral and right chiral complex helicity spinors are pure spinors (which are further constrained by a number of orthogonality and normalization conditions).

\subsection{10D polarized scattering equation }

The polarized scattering equations in D=10 is also doubled. The equations  imposed on left-chiral and right-chiral
helicity spinors corresponding to the scattered particles read
\begin{eqnarray}\label{pScEq=10DL=}
\sum\limits_{j\not=i} \frac {\lambda_{\alpha B j}{W}_{qj}^{{B}}{W}_{qi}^{{A}}}{\sigma_i-\sigma_j}=2\bar{\lambda}_{\alpha i}^{\; A} \; , \qquad \\
\label{pScEq=10DR=}
\sum\limits_{j\not=i} \frac { \lambda^{\alpha}_{B j}{W}_{\dot{q}j}^{{B}}{W}_{\dot{q}i}^{{A}}}{\sigma_i-\sigma_j}= 2\bar{\lambda}^{\alpha A}_i \; , \qquad
\end{eqnarray}
where the $4\times 8$ matrices ${W}_{qi}^{A}$ and ${W}_{\dot{q}i}^{A}$ obeys the purity conditions
 \begin{eqnarray}\label{WqAWqB=010D}
{W}_{qi}^{\; A} {W}_{qi}^{\; B}=0 \; , \qquad {W}_{\dot{q}i}^{A} {W}_{\dot{q}i}^{B}=0  \; . \qquad
\end{eqnarray}
Similar to 11D case, it is tempting to identify these with the blocks of the $i$-th internal frame matrices \eqref{bwwSO8}, \begin{eqnarray}\label{W=w}{W}_{qi}^{\; A}={w}_{qi}^{\; A}\; , \qquad {W}_{\dot{q}i}^{\; A}={w}_{\dot{q}i}^{\; A}\; . \qquad
\end{eqnarray}
We restrain ourselves from fixing rigidly such an identification at the present stage of the development of the formalism keeping in mind the  identification \eqref{W=w} but  keeping the notation of
${W}_{qi}^{\; A}$ and ${W}_{\dot{q}i}^{\; A}$ in the equations below.

Eqs. \eqref{pScEq=10DL=} and \eqref{pScEq=10DR=} are the counterparts of the 11D polarized scattering equations in the form of \eqref{ScEqli=lj}. To find the 10D counterpart of the polarized scattering equation in the form of Eq. \eqref{pScEq} we have to introduce two sets of constrained spinorial functions, $\lambda_{\alpha q}(\sigma)$ and $\lambda^{\alpha}_{\dot q}(\sigma)$, which obey
\begin{eqnarray}\label{Pmu=lsl}
2 \lambda_{{\alpha} {q}}(\sigma) \lambda_{{\beta} {q}}(\sigma)= {\sigma}{}^{{\mu} }_{{\alpha}{\beta}}P_{{\mu}}(\sigma)\; , \qquad P_{{\mu}}(\sigma) \delta_{{q}{p}}= \lambda_{{q}}(\sigma)\tilde{\sigma}{}_{{\mu} }\lambda_{{p}}(\sigma)\; , \qquad
\\ \label{Pmu=ltsl}
2 \lambda^{\alpha}_{\dot{q}}(\sigma) \lambda^{\beta}_{\dot{q}}(\sigma)= \tilde{\sigma}{}^{\mu\alpha\beta}P_{{\mu}}(\sigma)\; , \qquad
P_{{\mu}}(\sigma) \delta_{\dot{q}\dot{p}}= \lambda_{\dot{q}}(\sigma){\sigma}{}_{{\mu} }\lambda_{\dot{p}}(\sigma)\; , \qquad
\end{eqnarray}
where $P_{\mu}(\sigma)$ is the meromorphic 10-vector function \eqref{Pmu=}. One can check that the above constraints are satisfied if:
\\
i) the spinor functions are meromorphic functions of the form
\begin{eqnarray}\label{sol=pScEq10L} \lambda_{{\alpha} {q}}(\sigma )= \sum_{i=1}^n \sqrt{\rho^\#_i}\, \frac { v_{{\alpha}{A}i}^{\; -}{W}_{{q} i}^{\; {A}}(\sigma) }{\sigma-\sigma_i} = \sum_{i=1}^n \, \frac { \lambda_{{\alpha}{A}i}{W}_{{q} i}^{\; {A}}(\sigma) }{\sigma-\sigma_i}
\; ,
\qquad \\ \label{sol=pScEq10R} \lambda^{\alpha}_{\dot{q}}(\sigma )= \sum_{i=1}^n \sqrt{\rho^\#_i}\, \frac { v^{\alpha-}_{Ai}{W}_{\dot{q} i}^{\; {A}}(\sigma) }{\sigma-\sigma_i} = \sum_{i=1}^n \, \frac { \lambda^{\alpha}_{Ai}{W}_{\dot{q}i}^{\; {A}}(\sigma) }{\sigma-\sigma_i}
\; ,
\qquad
\end{eqnarray}
where
\begin{eqnarray}\label{Wi=WiO=10D}
 {W}_{q i}^{\; A} (\sigma)= {W}_{p i}^{\; A}\,  \tilde{\cal O}_{pq}(\sigma ) \; , \qquad {W}_{\dot q i}^{\; A} (\sigma)= {W}_{\dot p i}^{\; A}\,  \tilde{\cal O}_{\dot p\dot q}(\sigma ) \; \qquad
\end{eqnarray}
with $SO(8)$ valued matrices $\tilde{\cal O}_{pq}(\sigma )$ and $\tilde{\cal O}_{pq}(\sigma )$,
\begin{eqnarray}\label{cOcOT=I}
\tilde{\cal O}\tilde{\cal O}^T = {\bb I}_{8\times 8} \; ; \qquad
\end{eqnarray}
ii) the polarized scattering equations \eqref{pScEq=10DL=} and \eqref{pScEq=10DR=}  hold,
\\
iii)  ${W}_{p i}^{\; A}$ and ${W}_{\dot q i}^{\; A} $ obey \eqref{WqAWqB=010D}; this is automatic when \eqref{W=w} holds.

In terms of the meromorphic functions \eqref{sol=pScEq10L}
and \eqref{sol=pScEq10R} the polrized scattering equations \eqref{pScEq=10DL=} and \eqref{pScEq=10DR=} can be written in the form of
\begin{eqnarray}\label{pScEq=10DL}
\lambda_{\alpha q}(\sigma_i ) {W}_{qi}^{{A}}(\sigma_i ) =   2\sqrt{\rho^\#_i} \bar{v}{}_{\alpha i}^{\;{A}}=2\bar{\lambda}_{\alpha i}^{\; A} \; , \qquad \\
\label{pScEq=10DR}
\lambda^{\alpha}_{\dot q}(\sigma_i ) {W}_{{\dot q}i}^{{A}}(\sigma_i ) =   2\sqrt{\rho^\#_i} \bar{v}{}^{\alpha {A}}_i=2\bar{\lambda}^{\alpha A}_i \; . \qquad
\end{eqnarray}

Some comments are in order.
First of all, \eqref{WqAWqB=010D} and \eqref{Wi=WiO=10D} imply
 \begin{eqnarray}\label{WqAWqB=0=10D}
{W}_{qi}^{A}(\sigma) {W}_{qi}^{B} (\sigma)=0 \; , \qquad {W}_{\dot qi}^{A}(\sigma) {W}_{\dot qi}^{B} (\sigma)=0 \; .
\end{eqnarray}
Secondly,  the constraints \eqref{Pmu=lsl} and \eqref{Pmu=ltsl} can be solved by expressing the spinor fields in terms of spinor moving frame field and compensator field $\rho^\#(\sigma)$ by
\begin{eqnarray}\label{l=r12v10D}
\lambda_{\alpha q}(\sigma ) =   2\sqrt{\rho^\#(\sigma) } v_{\alpha q}^{\; -}(\sigma ) \; , \qquad \lambda^{\alpha}_{\dot q}(\sigma )  =   2\sqrt{\rho^\#(\sigma) } v^{\alpha-}_{\dot q}(\sigma ) \; . \qquad
\end{eqnarray}

This is the place to stress that, according to \eqref{sol=pScEq10L} and  \eqref{sol=pScEq10R}, $\lambda_{\alpha q}(\sigma ) $ and
$ \lambda^{\alpha}_{\dot q}(\sigma ) $ are complex so that $v_{\alpha q}^{\; -}(\sigma )$, $v^{\alpha-}_{\dot q}(\sigma ) $  and
$\rho^\#(\sigma) $ should be considered as complexifications of the spinor moving frame variables and densities used e.g. in \cite{BZ-str}. We refer to the last paragraph of sec. \ref{sec=lambda} for the discussion on such a complexification in 11D context.

Notice that  \eqref{l=r12v}, the 11D counterpart of  \eqref{l=r12v10D}, contains an additional $SO(16)$ matrix. Absence of the counterpart of this in \eqref{l=r12v10D} is explained by the fact that, if included, this should be $SO(8)$ valued matrix and the spinor frame variables which differ by $SO(8)$ gauge transformations  are considered to be identical  (see \cite{Galperin:1991gk,Delduc:1991ir} and \cite{Bandos:2017zap,Bandos:2017eof} for more details). In contrast, in 11D
the harmonics are identified modulo $SO(9)$ gauge symmetry while \eqref{Pmu=lGl} is invariant under a bigger $SO(16)$ group so that $SO(16)$ matrix enters naturally the general solution of \eqref{Pmu=lGl}.

\subsection{10D ambitwistor superstring and polarized scattering equation}
\label{sec=10Dambi}

The spinor moving frame or twistor-like formulation of the simplest ${\cal N}=1$  10D ambitwistor superstring, suitable for the description of 10D SYM and ${\cal N}=1$  $D=10$ supergravity  amplitudes,
can be based on the action quite similar to its 11D counterpart \eqref{SaTw=spf}
\cite{Bandos:2014lja}
\begin{eqnarray}\label{SaTw=spf10D}
S&=&\; \int_{{\cal W}^2}d^2\sigma
\lambda_{{\alpha}{q}} (\sigma ) \lambda_{{\beta}{q}}(\sigma )
\left( \bar{\partial}X^{{\alpha}{\beta}}(\sigma ) -i \bar{\partial}\theta ^{({\alpha}}\,  \theta^{{\beta})} (\sigma )\right)\;    \qquad \nonumber \\
&\equiv &\; \int_{{\cal W}^2}d^2\sigma
\rho^{\#}(\sigma )  v_{{\alpha}{q}}^{\; -}(\sigma )  v_{{\beta}{q}}^{\; -}(\sigma )
\left( \bar{\partial}X^{{\alpha}{\beta}}-i \bar{\partial}\theta ^{({\alpha}}\,  \theta^{{\beta})} \right)\;  .   \qquad
\end{eqnarray}
It is written in terms of constrained bosonic spinor functions obeying \eqref{Pmu=lsl}, 16-component fermionic spinor field  $\theta ^{{\alpha}}(\sigma)$ and  arbitrary symmetric spin tensor bosonic field
\begin{eqnarray}
\label{Xalbe=10D}
X^{{\alpha}{\beta}}(\sigma)=X^{{\beta}{\alpha}}(\sigma)\equiv
\frac {1}{16} \tilde{\sigma}_{{\mu}}{}^{{\alpha}{\beta}} X^{{\mu}}(\sigma)+ {1\over {2\cdot 16\cdot\, 5!}}
 Z^{{\mu}_1\ldots {\mu}_5}(\sigma)
 \tilde{\sigma}_{{\mu}_1\ldots {\mu}_5}{}^{{\alpha}{\beta}} \; .  \quad
\end{eqnarray}
The second form of the action \eqref{SaTw=spf10D}, which is obtained by substituting \eqref{l=r12v10D}, makes manifest the spinor moving frame nature of this twistor-like formulation of the 10D ambitwistor superstring.

Again, the properties of the spinor moving frame and spinorial functions  \eqref{l=r12v10D}, which are concenrated in \eqref{Pmu=lsl} and  \eqref{Pmu=ltsl}, guarantee that the arbitrary variation of $  Z^{{\mu}_1\ldots {\mu}_5}(\sigma)$ live the action invariant.  The gauge fixing condition for this local symmetry can be chosen to be
$  Z^{{\mu}_1\ldots {\mu}_5}(\sigma)=0$ so that
\begin{eqnarray}
\label{Xalbe=Xa10D}
X^{{\alpha}{\beta}}(\sigma)=
\frac {1}{16} \tilde{\sigma}_{{\mu}}^{{\alpha}{\beta}} X^{{\mu}}(\sigma)\; .  \qquad
\end{eqnarray}
 However, for our purposes it is more convenient  to treat the 10D ambitwistor superstring as a dynamical system in the enlarged superspace $\Sigma^{(136|32)}$ with $10+126=136$ bosonic coordinates
$(X^{{\mu}}, Z^{{\mu}_1\ldots {\mu}_5})$ and $16$ fermionic coordinates $\theta{}^{{\alpha}}$.

The constrained twistor form of the 10D ambitwistor superstring action and 10D generalization of the Penrose incidence relations look quite similar to their 11D counterparts (\ref{SaTw=sTw})--\eqref{eta=}:
\begin{eqnarray}\label{SaTw=sTw10D}
S_{10D}=\int_{{\cal W}^2}d^2\sigma \left( \lambda_{{\alpha}{q}}\,  \bar{\partial} {\mu}^{{\alpha}}_{{q}}-
\bar{\partial}\lambda_{{\alpha}{q }}\;  {\mu}^{{\alpha}}_{{q}} - i \bar{\partial}\eta_{{q}}\,\eta_{{q}}\right)
\;
\end{eqnarray}
and
\begin{eqnarray}\label{lambda=10D}
 \lambda_{{\alpha}{q}}(\sigma)= \sqrt{\rho^{\#}(\sigma)}  v_{{\alpha}{q}}^{\;-}(\sigma)\; ,  \qquad
\\
\label{mu=10D}
{\mu}^{{\alpha}}_{{q}}(\sigma)= X^{{\alpha}{\beta}} (\sigma)\lambda_{{\beta} {q}}(\sigma) - {i\over
2} \theta^{{\alpha}}(\sigma) \, \theta^{{\beta}}(\sigma) \lambda_{{\beta} {q}}(\sigma)  \; , \qquad \\
\label{eta=10D}
\eta_{{q}}(\sigma)= \theta^{{\beta}}(\sigma) \lambda_{{\beta} {q}}(\sigma) \; .  \qquad
\end{eqnarray}
The most noticed difference is the presence in \eqref{lambda=} the SO(16) matrix which has no counterpart in 10D equation \eqref{lambda=10D}.
As we have already commented, this is due to the fact that, if present in 10D, this should be $SO(8)$ valued matrix and $SO(8)$ is the fundamental gauge symmetry of the 10D spinor moving frame construction.

Eqs. (\ref{mu=10D}) and \eqref{eta=10D}
describe the general solution of $28$ constraints
\begin{eqnarray}
\label{SO(8)=} & {\mathbb{J}}_{{p}{q}}
:=
 2 \lambda_{{\alpha} [{p}}{\mu}_{{q}]}{}^{
{\alpha}} + i {\eta}_{{p}}{\eta}_{{q}} =0 \,  \qquad
\end{eqnarray}
which can be identified with generators of $SO(8)$ gauge symmetry in the Hamiltonian formalism.

The action (\ref{SaTw=sTw10D}) is  invariant under the gauge symmetry
\begin{eqnarray}
\label{vmu=gauge10D} \delta {\mu}^{{\alpha}}_{{q}}=   {1\over {32\cdot\, 5!}}
 \delta  Z^{{\nu}_1\ldots {\nu}_5}(\sigma)
 \tilde{\sigma}_{{\nu}_1\ldots {\nu}_5}{}^{{{\alpha}{\beta}}} \lambda_{{\beta} {q}}
\; ,  \quad
\end{eqnarray}
with arbitrary  $ \delta  Z^{{\nu}_1\ldots {\nu}_5}(\sigma)$, which
allows for the gauge fixing conditions reducing the general solution (\ref{mu=10D}) of the constraints to
\begin{eqnarray}
\label{mu=10D=}
{\mu}^{{\alpha}}_{{q}}:= \frac 1 {16} X^{{\nu}} \tilde{\sigma}_{{\nu}}^{{\alpha}{\beta}} \lambda_{{\beta} {q}} - {i\over 2} \theta^{{\alpha}} \, \theta^{{\beta}} \lambda_{{\beta} {q}}\; . \qquad
\end{eqnarray}

But for our purposes it is more convenient to do not fix this gauge symmetry. Then the only constraint restricting ${\mu}^{{\alpha}}_{{q}}(\sigma)$ is \eqref{SO(8)=}. Similarly to 11D case, we  can included this in the action with the Lagrange multiplier $ \bar{{\cal A}}{}^{{p}{q}}= \bar{{\cal A}}{}^{[pq]}$ playing the role of $SO(8)$ gauge field,
\begin{eqnarray}\label{SaTw=sTwL10}
S_{10D}=\int_{{\cal W}^2}d^2\sigma \left( \lambda_{{\alpha}{q}}\,  \bar{\partial} {\mu}^{{\alpha}}_{{q}}-
\bar{\partial}\lambda_{{\alpha}{q }}\;  {\mu}^{{\alpha}}_{{q}} - i \bar{\partial}\eta_{{q}}\,\eta_{{q}}\right)
+ \int_{{\cal W}^2}d^2\sigma \bar{{\cal A}}{}^{{p}{q}}\left( 2 \lambda_{{\alpha} [{p}}\mu^{
{\alpha}}_{{q}]} + i {\eta}_{[{p}}{\eta}_{{q}]}\right)
\;
\end{eqnarray}
and consider  the variables ${\mu}^{{\alpha}}_{{q}}$ as unconstrained.

Supersymmetry transformations leaving invariant the actions \eqref{SaTw=spf10D} and \eqref{SaTw=sTwL10} are
\begin{eqnarray}
\label{susy=X10D}  \delta_\epsilon X^{{\alpha}{\beta}} =i \theta^{({\alpha}}\epsilon^{{\beta})}\, , \qquad \delta_\epsilon  \theta^{{\alpha}}= \epsilon^{{\alpha}}\, , \qquad   \delta_\epsilon \lambda_{{\alpha} {q}}=0\, , \qquad
\end{eqnarray}
and
\begin{eqnarray}
\label{susy=sTw10D}  \delta_\epsilon \lambda_{{\alpha} {q}}=0\, , \qquad \delta_\epsilon {\mu}_{{q}}{}^{
{\alpha}} =- i \epsilon^{
{\alpha}} \eta_{{q}}\, , \qquad \delta_\epsilon \eta_{{q}}=  \epsilon^{
{\alpha}} \lambda_{\alpha {q}} \, . \qquad
\end{eqnarray}

Essentially in the same manner as in 11D case, Eq. \eqref{sol=pScEq10L} together with its fermionic superpartner,
\begin{eqnarray}\label{sol=pScEq10L=} \lambda_{{\alpha} {q}}(\sigma )= \sum_{i=1}^n \, \frac { \lambda_{{\alpha}{A}i}{W}_{{q} i}^{\; {A}}(\sigma) }{\sigma-\sigma_i}
\; ,
\qquad \end{eqnarray}
and
\begin{eqnarray} \label{sol=ScEqF=10D}
\eta_{{q}}(\sigma )=  \sum_{i=1}^n  \frac { \eta_{Ai}{W}_{{q} i}^{\; A} (\sigma)}{\sigma-\sigma_i}\;  \qquad
\end{eqnarray}
with ${W}_{{q} i}^{\; A} (\sigma)$ from Eq. \eqref{Wi=WiO=10D} and $\eta_{Ai}=\eta_{q i}\bar{w}_{qA \, i}$, can be obtained as the solutions of saddle point equations for the path integral with the measure defined by the ambitwistor superstring action \eqref{SaTw=sTwL10} and the suitable vertex operator,
\begin{eqnarray}\label{V=int10D}
V=  \int d^2\sigma_i \delta (k_i\cdot P(\sigma_i)) \frak{W} \exp \left(2i\mu^{{\alpha}}_{{q}} (\sigma_i)
\lambda_{{\alpha}Ai} W_{{q}i}^A (\sigma_i)+ 2\eta_{{q}}(\sigma_i)
 \eta_{Ai} W_{{q}i}^A(\sigma_i)
\right)
\;   \qquad
\end{eqnarray}
({\it cf.} \eqref{V=int} and discussion around).

What is specific for 10D is the problem of how to obtain the corresponding equation for $\lambda^{\alpha}_{\dot{q}}(\sigma )$ which do not enter (explicitly) the ambitwistor superstring action,
\begin{eqnarray}
\label{sol=pScEq10R=} \lambda^{\alpha}_{\dot{q}}(\sigma )= \sum_{i=1}^n \sqrt{\rho^\#_i}\, \frac { v^{\alpha-}_{Ai}{W}_{\dot{q} i}^{\; {A}}(\sigma) }{\sigma-\sigma_i} = \sum_{i=1}^n \, \frac { \lambda^{\alpha}_{Ai}{W}_{\dot{q}i}^{\; {A}}(\sigma) }{\sigma-\sigma_i}
\; .
\qquad
\end{eqnarray}
Although it is intuitively clear that this should be the case due to that $\lambda^{\alpha}_{\dot{q}}(\sigma )$ and
$ \lambda_{{\alpha} {q}}(\sigma )$ are different forms of the square root of the meromorphic ten-vector function \eqref{Pmu=} (in the sense of
constraints \eqref{Pmu=lsl} and \eqref{Pmu=ltsl}) the understanding of the spinor moving frame nature of both the spinorial functions and helicity spinors helps to provide a more explicit arguments in favour of this. To this end, besides the generic statement that $\lambda^{\alpha}_{\dot{q}}(\sigma )$ and
$ \lambda_{{\alpha} {q}}(\sigma )$ represent the same element of the coset space ${SO(1,9) \over SO(8)}$ isomorphic to ${\bb S}^{8}\otimes {\bb R}_+$, one can use the fact that thier derivatives are expressed in terms of the same Cartan forms (see Appendix \ref{CartanF}) or a special parametrization of spinor frame variables found in \cite{Bandos:2017eof} in which $\lambda^{\alpha}_{\dot{q}}(\sigma )$ and
$ \lambda_{{\alpha} {q}}(\sigma )$, as well as
$P_\mu(\sigma)$,  are expressed in terms of the same parameter functions ${\bb K}^{=I}(\sigma)$ and   $\rho^\#(\sigma)$ (see Eqs. (7.42)-(7.50) in \cite{Bandos:2017eof}). Then  \eqref{sol=pScEq10L=} and \eqref{sol=pScEq10R=} provide equivalent expressions for these parameter functions.

\subsection{10D supersymmetry generator and supersymmetric invariants}

From \eqref{susy=sTw10D} it is easy to restore the form of the ${\cal N}=1$ supersymmetry generator $Q_\alpha$ which obeys the superalgebra
$\{ Q_\alpha ,  Q_\beta \}= 4\lambda_{\alpha q}\lambda_{\beta q}=8\lambda_{(\alpha | A}\lambda_{|\beta)}^{\; A}$. Its realization on the variables of $i$-th of scattered particles  reads
\begin{eqnarray}\label{Qali=10D}
 Q_{\alpha i} =4\lambda_{\alpha i}^{\;A}\eta_{Ai}+
\lambda_{\alpha Ai} \frac {\partial}{\partial  \eta_{Ai}}\; , \qquad \{ Q_{\alpha i} ,  Q_{\beta j} \}=8\delta_{ij} \lambda_{(\alpha | Ai}\lambda_{|\beta)i}^{\; A}= 2\delta_{ij}k_{\mu i} \sigma^{\mu}_{\alpha \beta}\; .
\end{eqnarray}
The complete supersymmetry generator given by the sum of the partial generators
\begin{eqnarray}\label{Qal=SQali=10D}
 Q_{\alpha } =\sum\limits_i Q_{\alpha i} =\sum\limits_i \left(4\lambda_{\alpha i}^{\;A}\eta_{Ai}+
\lambda_{\alpha Ai} \frac {\partial}{\partial  \eta_{Ai}}\right)\;
\end{eqnarray}
is nilpotent, $ \{ Q_{\alpha} ,  Q_{\beta} \}=0$,  due to the momentum conservation.

The supersymmetric invariant found in \cite{Geyer:2019ayz} is $e^{{\bb F}}$ with
\begin{eqnarray}\label{F=10D}
{\bb F}&=& 2\sum\limits_i   \sum\limits_j  \frac {W^{{A}}_{{q}j}W^{{B}}_{{q}i}}{\sigma_j-\sigma_i}\eta_{{A}j}\eta_{{B}i}
\; .
\end{eqnarray}
The  superamplitudes of 10D SYM are then essentially described by Eqs. \eqref{S-mat=11D}--\eqref{det'=} where the reduced determinant $det'$ is replaced by reduced Pfafian $Pf'$ and all the variables are considered to be ten dimensional.

The generalizations of supersymmetric invariants to type II cases is straightforward  \cite{Geyer:2019ayz}. As far as the derivation of  the basic equation for spinorial function  is concerned, the generalization of our discussion in sec. \ref{sec=10Dambi} is straightforward for IIB case while type IIA case seems to be problematic. The issue can be seen from the Lagrangian 1-form associated to the 10D type IIA ambitwistor superstring action,
$$\lambda_{\alpha q} \lambda_{\beta q}dX^{\alpha\beta} - i \lambda_{\alpha q} \lambda_{\beta q}d\theta^{\alpha}_1\theta^{\beta }_1- i \lambda^{\alpha}_{\dot q} \lambda^{\beta}_{\dot{q}}d\theta_{\alpha 2}\theta_{\beta  2}$$ with $X^{\alpha\beta}=\frac 1 {16} X^\mu \tilde{\sigma}_\mu^{\alpha\beta}$. It is gauge equivalent to a Lagrangian form in an enlarged superspace with $10+126$ bosonic  coordinates described by an arbitrary symmetric spin-tensor $X^{\alpha\beta}=X^{\beta\alpha}$. However,  supersymmetry transformations living invariant such a generalization of the Lagrangian form,
$$
\delta X^{\alpha\beta}=i\theta^{(\alpha}_1\epsilon^{\beta )}_1 + \frac {i}{16}\tilde{\sigma}_\mu^{\alpha\beta}\tilde{\sigma}^{\mu\gamma\delta}\theta_{\gamma 2}\epsilon_{\delta 2}\; , \qquad \delta \theta^{\alpha}_1=\epsilon^{\alpha}_1\; , \qquad  \delta\theta_{\alpha 2}=\epsilon_{\alpha 2}\;,  \qquad
$$
are quite asymmetric and it is not clear whether it is possible to introduce a convenient supertwistor variables providing the basis of (constrained) Darboux coordinates  for this Lagrangian form. Thus it seems that in  type IIA case the shortcut through the enlarged superspace does not work and to obtain equation for the bosonic spinor functions one has to deal with the action containing supertwistor variables  restricted by additional constraints similar to \eqref{moreII=}.

\section{Conclusion and discussion}

In this paper we have revisited the formalism of the 11D polarized scattering equations of \cite{Geyer:2019ayz} from the point of view of spinor frame approach different applications of which to the description of 11D and 10D amplitudes were searched for in \cite{Bandos:2016tsm,Bandos:2017zap,Bandos:2017eof}. In particular, we have addressed the problem of rigorous derivation of the equations for spinorial meromorphic function $\lambda_{\underline{\alpha}\underline{q}}(\sigma)$ and its fermionic superpartner
$\eta_{\underline{q}}(\sigma)$ from the (spinor moving  frame formulation) of 11D ambitwistor superstring  \cite{Bandos:2014lja}. We have shown that, to this end, the (gauge equivalent) formulation of ambitwistor  superstring as dynamical system in an enlarged 11D superspace
$\Sigma^{(528|32)}$ with additional tensor central charge coordinates is very useful.

The polarized scattering equation can be written in two equivalent forms: as Eq. \eqref{pScEq} for the spinor function on the Riemann sphere, and as Eq. \eqref{ScEqli=lj} imposed on the scattering data.
We have  found the fermionic superpartner of the polarized scattering equation
 \eqref{ScEqli=lj}. We call this {\it spolarized scattering equation}. It happens to be an equation imposed on the supergravity amplitude, Eq. \eqref{feq-S-mat=11D}, rather then on the scattering data.

We have also revised  the 10D polarized scattering equation formalism and its 10D ambitwistor superstring origin with the use of spinor frame method. In this case a counterpart of hidden $SO(16)$ symmetry of the 11D ambitwistor superstring does not appear, being replaced by $SO(8)$ symmetry characteristic for the spinor frame formalism. However, similarly to 11D case, the treatment of   the ambitwistor superstring as a dynamical system in 10D superspace enlarged by $126$ directions parametrized by tensorial central charge coordinates is also useful to obtain the basic equations for the spinor functions.

An interesting direction for future study  is to apply the spinor frame approach to the construction of
11D and 10D generalization of the 6D rational map and symplectic Grassmannians approach  \cite{Heydeman:2017yww,Cachazo:2018hqa,Heydeman:2018dje,Schwarz:2019aat}. Its relation to the 6D polarized scattering equation approach of \cite{Geyer:2018xgb} was discussed in very recent \cite{Schwarz:2019aat}.

The rational maps approach introduces a scattering map
\be\label{cP=pi}
{\cal P}_{\underline{\mu}} (\sigma) = \sum\limits_{i=1}^n k_{\underline{\mu}i}\prod_{j\not=i}(\sigma-\sigma_j)
\ee
 instead of the ambitwistor superstring momentum function \eqref{Pmu=}. Clearly
 ${\cal P}_{\underline{\mu}} (\sigma) =  P_{\underline{\mu}} (\sigma)\prod_{j=1}^n (\sigma-\sigma_j)$ and the scattering equation can be also obtained from the light-likeness condition of the scattering map
 \be\label{cP2=0}
{\cal P}_{\underline{\mu}} (\sigma){\cal P}^{\underline{\mu}} (\sigma) = 0\; .
\ee
Extrapolating the 6d results of \cite{Cachazo:2018hqa,Schwarz:2019aat} one might expect that in 11D spacetime this can be solved  in a manner similar to \eqref{P=lGl},
 \be\label{cP2=rGr}
{\cal P}_{\underline{\mu}} (\sigma) {\Gamma}{}^{\underline{\mu} }_{\underline{\alpha}\underline{\beta}} =
2 \rho_{\underline{\alpha} \underline{q}}(\sigma)\rho_{\underline{\beta} \underline{q}}(\sigma) \; , \qquad
{\cal P}_{\underline{\mu}} (\sigma)\delta_{\underline{q}\underline{p}}=  \rho_{ \underline{q}}(\sigma )\tilde{\Gamma}{}_{\underline{\mu} }\rho_{\underline{p}}(\sigma )\; ,
\ee
but in terms of rational spinor map $\rho_{\underline{\alpha} \underline{q}}(\sigma)$ (instead of meromorphic function
$\lambda_{\underline{\alpha} \underline{q}}(\sigma)$ \eqref{sol=pScEq1}) which for even $n=2m+2$  has the form
\be\label{r=rksk}
\rho_{\underline{\alpha} \underline{q}}(\sigma)= \sum\limits_{k=0}^m \rho_{\underline{\alpha} \underline{q},k}\, \sigma^k\; .
\ee

Of course, in distinction to 6d and 4d cases, the 11D equations  \eqref{cP2=rGr} (and their 10D counterparts) impose strong constraints on $\rho_{\underline{\alpha} \underline{q}}(\sigma)$ so that their consistency with \eqref{r=rksk} has to be checked.
We leave this problem for future work and conclude here by observation that, if this consistency holds, the relation between
 coefficients of the rational maps and the helicity spinors, encoding the scattering data through  \eqref{kI=lGl} and
\eqref{UG=vUv} with \eqref{l=sqrv}, should be  described  by\footnote{To find this one notices that Eq. \eqref{cP=pi} \cite{Cachazo:2018hqa} implies $k_{\underline{\mu}i}=
\frac 1 {2\pi i} \oint\limits_{|z-\sigma_i|=\epsilon}dz   \frac {{\cal P}_{\underline{\mu}} (z)}{\prod_{j}(z-\sigma_j)}$ and uses
\eqref{kI=lGl} and \eqref{cP2=rGr}.}
\begin{eqnarray}\label{li=rsi}
\lambda_{\underline{\alpha} \underline{q}i}= \;\frac
{ \rho_{\underline{\alpha} \underline{p}}(\sigma_i ) \tilde{{\cal S}}_{\underline{p}\underline{q}} (\sigma_i )}{\sqrt{\prod_{j\not=i}(\sigma_i-\sigma_j)}}
 \qquad
\end{eqnarray}
with some $SO(16)$ valued matrix function $\tilde{{\cal S}}_{\underline{p}\underline{q}} (\sigma )$, $\; \tilde{{\cal S}}\tilde{{\cal S}}^T={\bb I}_{16\times 16}$ and $\rho_{\underline{\alpha} \underline{p}}(\sigma_i ) $ given in \eqref{r=rksk}.

\bigskip

\subsection*{Notice added.}

When this paper have been finished and ready for sending to the arXive, the article \cite{Berkovits:2019bbx} appear on the net. There another supertwistor formulation of ambitwistor superstring was considered, quantized in light cone gauge   and compared with the light cone gauge description of the RNS type formulation of the ambitwistor superstring \cite{Mason:2013sva}. The light cone gauge scattering amplitudes have been also discussed in \cite{Berkovits:2019bbx}.

The supertwistors used in \cite{Berkovits:2019bbx} were introduced in \cite{Berkovits:1990yc} in the context of massless superparticle model (see also \cite{Witten:1985nt}). The components of that supertwistor  are an unconstrained 16-component bosonic spinor
$\lambda^\alpha$, canonically conjugate to it 16-component bosonic spinor $w_\alpha$, and fermionic 10-vector $\psi^\mu$. Thus, on one hand, the fermionic variables of this alternative  supertwistor formulation of the ambitwistor string are  RNS-like and, on the other hand,  it uses essentially the representation of a light-like vector function as a bilinear of single unconstrained bosonic spinor, $P_\mu (\sigma)= \lambda^\alpha (\sigma)\sigma_{\mu\alpha\beta} \lambda^\beta (\sigma)$. This is valid  due to the specific  identity  for D=10 $\sigma$-matrices (having its counterparts also in $D=3,4,6$) and,
in distinction to our spinor moving frame related constrained supertwistor approach,  do not allow for a straightforward generalization to 11D case.

\bigskip

\subsection*{Acknowledgements}
This work was supported in part by the Spanish MINECO/FEDER (ERDF EU)  grant PGC2018-095205-B-I00, by the Basque Government Grant IT-979-16, and by the Basque Country University program UFI 11/55.

\bigskip

\appendix
\renewcommand{\theequation}{A.\arabic{equation}}
\section{Some properties of 11D spinor frame variables  and helicity spinors}

In our mostly minus metric conventions the 11D Dirac matrices $\gamma_{\underline \mu}{}_{\underline \alpha}{}^{\underline \beta}$ obeying
$$
\gamma_{\underline \mu}\gamma_{\underline \nu}+ \gamma_{\underline \nu} \gamma_{\underline \mu}= \eta_{\underline \mu\underline \nu}\; {\bb I}_{32\times 32} = {\rm diag} (+1,\underbrace{-1,...,-1}_{10}) \; {\bb I}_{32\times 32}\;
$$
 are imaginary. The charge conjugation matrix $C^{\underline{\alpha}\underline \beta}$  and its inverse $C_{\underline{\alpha}\underline \beta}$ are imaginary as well. We use mainly the matrices with both upper and with both lower indices
 $$
\Gamma_{\underline \mu}{}_{\underline{\alpha}\underline \beta}:= \gamma_{\underline \mu}{}_{\underline \alpha}{}^{\underline \gamma} C{}_{\underline \gamma\beta}= \Gamma_{\underline \mu}{}_{\underline \beta\underline{\alpha}}\; , \qquad
\tilde{\Gamma}_{\underline \mu}{}^{\underline{\alpha}\underline \beta}:= C^{\underline{\alpha}\underline \gamma}\gamma_{\underline \mu}{}_{\underline \gamma}{}^{\underline \beta}
=\tilde{\Gamma}_{\underline \mu}{}^{\underline \beta \underline{\alpha}}
$$
which are real, symmetric and, by construction, obey \eqref{G=CG}.

\subsection{Spinor frame and vector frame variables (Lorentz harmonics) in D=11}

Interrelations between D=11 vector frame and 11D spinor frame variables are described by
\begin{eqnarray}\label{u==v-v-11D}
 & u_{\underline \mu}^= \Gamma^{\underline \mu}_{\underline\alpha\underline \beta}= 2v_{\underline\alpha \underline q}^{\; -} v_{\underline\beta \underline q}^{\; -}  \;  & ,  \qquad
 v^-_{\underline{q}} \tilde{\Gamma}_{\underline \mu }v^-_{\underline{p}}= u_{\underline \mu}^= \delta_{\underline{q}\underline{p}}\; , \qquad
\\
\label{v+v+=u++=11D}
& u_{\underline \mu}^{\# }{\Gamma}^{\underline \mu}_{ {\alpha} {\beta}} = 2 v_{\underline{\alpha}\underline{q}}^{\; +}v_{\underline{\beta}\underline{q}}^{\; +}\; & , \qquad v_{\underline{q}}^+ \tilde{\Gamma}_{ \underline{\mu}} v_{\underline{p}}^+ = \; u_{\underline {\mu}}^{\# } \delta_{\underline{q}\underline{p}}\; ,  \qquad   \\
\label{uIs=v+v-=11D}
&  u_{\underline {\mu}}^{I} {\Gamma}^{\underline \mu}_{\underline\alpha\underline\beta}=  2 v_{( \underline{\alpha}|\underline{q} }{}^- \gamma^I_{\underline{q}\underline{p}}v_{|\underline{\beta})\underline{p}}{}^{+} \; &, \qquad  v_{\underline{q}}^- \tilde {\Gamma}_{\underline {}} v_{\underline{p}}^+=u_{\underline \mu}^{I} \gamma^I_{\underline{q}\underline{p}}\; , \qquad
\end{eqnarray}
where $\underline{q},\underline{p}=1,...,16$ are spinor indices of SO(9) and
$\gamma^I_{\underline{q}\underline{p}}=\gamma^I_{\underline{p}\underline{q}}$ are SO(9) gamma matrices.

In addition to the above  spinor frame variables \eqref{harmV=10D} we have also used the elements of
the inverse of the spinor moving frame matrix
\begin{eqnarray}\label{harmV-1=D}
V_{(\underline{\beta})}^{\;\;\; \underline{\alpha}}= \left(\begin{matrix}  v^{+\underline{\alpha}}_{\underline{q}}
 \cr  v^{-\underline{\alpha}}_{\underline{q}} \end{matrix} \right) \quad \in \quad Spin(1,10)
 \;  \qquad
\end{eqnarray}
the blocks of which obey  $V_{\underline{\alpha}}^{(\underline{\beta})}V_{(\underline{\beta})}{}^{\underline{\gamma}} :=  v_{\underline{\alpha}{\underline{q}}}^{\; -} v^{+\underline{\gamma}}_{\underline{q}}
+ v_{\underline{\alpha}{\underline{q}}}^{\; +} v^{-\underline{\gamma}}_{\underline{q}}
=
\delta_{\underline{\alpha}}^{\;\;\underline{\gamma}}$ and
\begin{eqnarray}\label{v-qv+p=}
&
v^{+\underline{\alpha}}_{\underline{q}} v_{\underline{\alpha} \underline{p}}^{\; -} = \delta_{\underline{q}\underline{p}}
 \; ,  \qquad & v^{+\underline{\alpha}}_{\underline{q}}  v_{\underline{\alpha} \underline{p}}^{\; +}=0\;  , \qquad
 \nonumber  \\
 &
 v^{-\underline{\alpha}}_{\underline{q}}   v_{\underline{\alpha} \underline{p}}^{\; -}=0\;  , \qquad & v^{-\underline{\alpha}}_{\underline{q}}   v_{\underline{\alpha} \underline{p}}^{\; +}=\delta_{\underline{q}\underline{p}}\;  . \qquad
\end{eqnarray}

In D=11 the elements of the inverse spinor frame matrix can be constructed from the elements of  (\ref{harmV=11D}) with the use of charge conjugation matrix  \begin{eqnarray}
\label{V-1=CV-A} D=11\; : \qquad  v_{\underline q}^{\pm \underline  \alpha}= \pm  i C^{\underline \alpha\underline \beta}v^{\; \pm}_{\underline \beta \underline q}\, .
\qquad
 \end{eqnarray}
The relations between $v_{\underline q}^{- \underline  \alpha}$ and $v^{\; -}_{\underline \beta \underline q}\,$,
$\;v_{\underline q}^{- \underline  \alpha}= -  i C^{\alpha\beta}v^{\; -}_{\underline \beta \underline q}$ coincide with our conventions for rising and lowering the 11D Majorana spinor indices which imply, e.g.
\begin{eqnarray}
\label{lalU=ClalD}  \lambda_{\underline q}^{\underline  \alpha}= - i C^{\underline \alpha\underline \beta}\lambda_{\underline \beta \underline q}\, , \qquad \lambda_{\underline  \alpha \underline q}=  i C_{\underline \alpha\underline \beta}\lambda_{\underline q}^{\underline  \alpha}\, ,
\qquad
 \end{eqnarray}
and $\Gamma_{\underline \mu \underline \alpha \underline \beta}= i C_{ \underline \alpha \underline \gamma}iC_{ \underline \beta \underline \delta }\tilde{\Gamma}_{\underline \mu}{}^{\underline \gamma\underline \delta }=
C_{ \underline \alpha \underline \gamma}\tilde{\Gamma}_{\underline \mu}{}^{\underline \gamma\underline \delta }C_{ \underline \delta  \underline \beta }
$, while the sign in the relation for complementary elements of the spinor frame, $v_{\underline q}^{+ \underline  \alpha}=  i C^{\underline \alpha\underline \beta}v^{\; +}_{\underline \beta \underline q}$, is opposite.

Notice that \eqref{lalU=ClalD} and \eqref{V-1=CV-A} implies that Eqs. \eqref{l=sqrv} , \eqref{lA:==11D} and \eqref{l=r12v} are valid also for the spinors with upper indices, while e.g. the upper-index version of \eqref{UG=vUv} has the opposite sign,
\begin{eqnarray}\label{UG=-vUv} \tilde{U}\!\!\!\!/{}^{\underline{\alpha}\underline{\beta}}:= U_{\underline{\mu}}\tilde{\Gamma}^{\underline{\mu}\underline{\alpha}\underline{\beta}}= -2
{v}^{-(\underline{\alpha}}_{\underline{q}} \gamma^{\underline{I}}_{\underline{q}\underline{p}}v^{+\underline{\beta})}_{\underline{p}} U^{\underline{I}}_i \; .
\;  \qquad
\end{eqnarray}

The different signs for $v^+$ and $v^-$ in \eqref{V-1=CV-A} are also reflected in the following
 consequences of the above constraints:
$$
(v^-_{\underline{q}}\tilde{\Gamma}_{\underline{\mu}})^{\underline{\alpha}}= u_\mu^= v^{+ \underline{\alpha} }_{\underline{q}} + u_{\underline{\mu}}^{\underline{I}}\gamma^{\underline{I}}_{\underline{q}\underline{p}} v^{-\underline{\alpha}} _{\underline{p}}\; , \qquad (v^-_{\underline{q}}\Gamma_{\underline{\mu}})_{\underline{\alpha}}= u_\mu^= v_{\underline{\alpha}\underline{q}}^+ - u_\mu^I \gamma^I_{\underline{q}\underline{p}}v_{\underline{\alpha} \underline{p}}^-\; , \qquad
$$
which imply
\bea\label{v-Gmnv-=}
v^-_{\underline{q}}\Gamma_{\underline{\mu}\underline{\nu}}v^-_{\underline{p}}= 2u_{[\underline{\mu}}^= u_{\underline{\nu}]}^{\underline{I}} \gamma^{\underline{I}}_{\underline{q}\underline{p}}\; .
\eea

\subsection{Internal frame variables/internal harmonics }

The internal frame variables or $SO(9)/[SO(2)\times SO(7)]$ harmonics can be described \cite{Bandos:2017zap} by  complex $16\times 8$ matrices $\bar{w}_{\underline{q}\underline{A}}= ({w}_{\underline{q}}{}^{\underline{A}})^*$ \eqref{harm=SO9} obeying \eqref{wbw+cc=1}, \eqref{bww=1} as well as
\bea\label{bwgIbw=UI}
U\!\!\!\!/{}_{\underline{q}\underline{p}}&=& 2 \bar{w}_{\underline{q}\underline{A}}\bar{w}_{\underline{p}\underline{A}}\; , \qquad
\bar{w}_{\underline{q}\underline{A}}\gamma^{\underline{I}}_{\underline{q}\underline{p}}\bar{w}_{\underline{p}\underline{B}}=
U^{\underline{I}}\delta_{\underline{A}\underline{B}}
\; , \qquad \\
\label{wgIw=bUI}
\overline{U}\!\!\!\!/{}_{\underline{q}\underline{p}}&=& 2 {w}_{\underline{q}}{}^{\underline{A}}{w}_{\underline{p}}{}^{\underline{A}}\; , \qquad
{w}_{\underline{q}}{}^{\underline{A}}\gamma^{\underline{I}}_{\underline{q}\underline{p}}{w}_{\underline{p}}{}^{\underline{B}}=
\bar{U}^{\underline{I}}\delta^{\underline{A}\underline{B}}
\; , \qquad \\
\label{bwgIbw=UIK}
U\!\!\!\!/{}^{\hat{K}}_{\underline{q}\underline{p}}&=& 2 {w}_{(\underline{q}|}{}^{\underline{A}} (\tau^{\hat{K}}){}_{\underline{A}}{}^{\underline{B}}\bar{w}_{|\underline{p})\underline{B}} \; , \qquad
\bar{w}_{\underline{q}\underline{A}}\gamma^{\underline{I}}_{\underline{q}\underline{p}}{w}_{\underline{p}}{}^{\underline{B}}=
U_{\underline{I}}{}^{\hat{K}} (\tau^{\hat{K}})_{\underline{A}}{}^{\underline{B}}
\; , \qquad \eea
where $(\tau^{\hat{K}}){}_{\underline{A}}{}^{\underline{B}}$ are $SO(7)$ Dirac matrices, $\hat{J}, \hat{K}=1,...,7$ and the vectors $U_{\underline{I}}$,  $\bar{U}_{\underline{I}}=({U}_{\underline{I}})^*$, $U_{\underline{I}}{}^{\hat{J}}$ form the $SO(9)$ valued matrix
\bea\label{UinSO9}
\left(U_{\underline{I}}{}^{\hat{J}} , \frac{1}{2} \left(U_{\underline{I}}+  \bar{U}_{\underline{I}} \right) , \frac{1}{2i} \left(U_{\underline{I}}- \bar{U}_{\underline{I}} \right)   \right)\; \in \; SO(9)
\;  \qquad \eea
which describes the vector internal frame.  The condition \eqref{UinSO9} implies
\bea\label{UbU=2}
U_{\underline{I}}U_{\underline{I}}=0\; , \qquad {U}_{\underline{I}} \bar{U}_{\underline{I}} =2 \; , \qquad \bar{U}_{\underline{I}} \bar{U}_{\underline{I}} =0\; , \qquad \nonumber \\  U_{\underline{I}} U_{\underline{I}}{}^{\hat{J}} =0 \; , \qquad \bar{U}_{\underline{I}}  U_{\underline{I}}{}^{\hat{J}} =0 \; , \qquad U_{\underline{I}}{}^{\hat{J}} U_{\underline{I}}{}^{\hat{K}} =\delta^{\hat{J}\hat{K}}\; . \qquad
\eea

Using the above properties of the internal harmonics and  \eqref{v-Gmnv-=} we can obtain Eq. \eqref{v-AG2v-B=kU},
\be\label{lAG2lB=11D}
\lambda_{\underline{A}}\Gamma_{\underline{\mu}\underline{\nu}}\lambda_{\underline{B}}=\rho^{\#} v^{-}_{\underline{A}}\Gamma_{\underline{\mu}\underline{\nu}}v^{-}_{\underline{B}}= 2 k_{[\mu} U_{\nu]}\delta_{\underline{A}\underline{B}}\; .
\ee

\section{An interesting nilpotent matrix}
\renewcommand{\theequation}{B.\arabic{equation}}
\setcounter{equation}0

Here we present an interesting $16\times 16$ nilpotent matrix which might happen to be useful in further development of the formalism.

The scattering equation in the form of (\ref{kiPsi=0}) implies
\begin{eqnarray}\label{Pki+kiP=0}
 \{P\!\!\!\!/ (\sigma_i) , k_i\!\!\!\!/ \}=
 P\!\!\!\!/_{\underline{\alpha}\underline{\gamma}}(\sigma_i)\tilde{k_i}\!\!\!\!/^{\underline{\gamma}\underline{\beta}} + k_i\!\!\!\!/_{\,\underline{\alpha}\underline{\gamma}}\tilde{P}\!\!\!\!/^{\underline{\gamma}\underline{\beta}}(\sigma_i)=  0 . \qquad
\end{eqnarray}
Using (\ref{kI=lGl}) and  (\ref{P=lGl}) this equation can be written in the equivalent form of
\begin{eqnarray}\label{W(vv-vv)=0}
0= W_{\underline{q}\underline{p}i}\left( v_{\underline{\alpha} \underline{q} i}^{\; -} v_{\underline{p}}^{-\underline{\beta}}(\sigma_i) - v_{\underline{q} i}^{-\underline{\beta}} v_{\underline{\alpha} \underline{p} }^{\; -} (\sigma_i)\right)  \qquad
\end{eqnarray}
where
\begin{eqnarray}\label{Wqpi=}
& W_{\underline{q}\underline{p}i}\frac {1}{\sqrt{\rho^{\#}_i\rho^{\#}(\sigma_i) }} = v_{\underline{\gamma} \underline{q} i}^{\; -} v_{\underline{p}}^{-\underline{\gamma}}(\sigma_i)\equiv  - v_{\underline{\gamma} \underline{p} }^{\; -} (\sigma_i) v_{\underline{q} i}^{-\underline{\gamma}}  \; . \qquad
\end{eqnarray}
Contracting  (\ref{W(vv-vv)=0}) with $ v_{\underline{\beta} \underline{q} i}^{\; -} $ and $ v_{\underline{\beta} \underline{q} }^{\; -}(\sigma_i) $ we find  nilpotency conditions for the
$W_{qpi}$ matrix,
\begin{eqnarray}\label{WWT=0}
W_{\underline{q}\underline{p}i} W_{\underline{q}\underline{p}'i} = 0 \; , \qquad W_{\underline{q}\underline{p}i} W_{\underline{q}'\underline{p}i} = 0 \; . \qquad
\end{eqnarray}
It is not difficult to check that these nilpotency conditions are equivalent to the scattering equation (\ref{kiPsi=0}).

Using \eqref{sol=pScEq1} we can write the above nilpotent matrix (\ref{Wqpi=}) in  the form
$$W_{\underline{q}\underline{p}i}=- \sum_{j=1, j\not=i}^n \, \lambda^{\underline{\alpha} }_{\underline{p}i} \; \frac  1 {\sigma_i-\sigma_j}{ \lambda_{\underline{\alpha}\underline{A}j}{W}_{\underline{q} j}^{\; \underline{A}} }\; . $$

\renewcommand{\theequation}{C.\arabic{equation}}
\section{Some properties of 10D  spinor frame variables and helicity spinors}

10D vector frame and spinor frame variabes  are related by
\begin{eqnarray}\label{u==v-v-10D}
   v^-_{{q}} \tilde{\sigma}_{a}v^-_{{p}}= u_a^= \delta_{{q}{p}} &\; , & \qquad
u_a^= \sigma^a_{\alpha\beta}= 2v_{\alpha q}{}^- v_{\beta q}{}^-  \; ,   \qquad  \\ \label{u==v-v-=10D}  v^-_{\dot{q}} {\sigma}_{a}v^-_{\dot{p}}= u_a^= \delta_{\dot{q}\dot{p}}
 & \; ,  & \qquad u_a^= \tilde{\sigma}{}^{a\, \alpha\beta}= 2v^{-\alpha}_{\dot q}   v^{-\beta}_{\dot q} \; , \qquad
\\
\label{v+v+=u++10D}
 v_{\dot{q}}^+ \tilde{\sigma}_{ {a}} v_{\dot{p}}^+ = \; u_{ {a}}^{\# } \delta_{\dot{q}\dot{p}} & \; , & \qquad  u_{ {a}}^{\# }  {\sigma}^{ {a}}_{ {\alpha} {\beta}}= 2 v_{{\alpha}\dot{q}}{}^{+}v_{{\beta}\dot{q}}{}^{+}\; , \qquad \\
\label{v+v+=u++10D-}
 v_{{q}}^+ {\sigma}_{ {a}} v_{{p}}^+ = \; u_{ {a}}^{\# } \delta_{{q}{p}} & \; , & \qquad  u_{ {a}}^{\# }  \tilde{\sigma}^{ {a} {\alpha} {\beta}} = 2 v_{{q}}^{+ {\alpha}}v_{{q}}^{+}{}^{ {\beta}} \; , \qquad
\\ \label{uIs=v+v-10D}
 v_{{q}}^- \tilde {\sigma}_{ {a}} v_{\dot{p}}^+=u_{ {a}}^{I} \gamma^I_{q\dot{p}} &\; , &\qquad
  u_{ {a}}^{I} {\sigma}^{a}_{\alpha\beta}= 2 v_{( {\alpha}|{q} }{}^- \gamma^I_{q\dot{q}}v_{|{\beta})\dot{q}}{}^{+} \; , \quad\\ \label{uIs=v-v+10D}
 v_{\dot q}^- {\sigma}_{ {a}} v_{{p}}^+ = - u_{ {a}}^{I} \gamma^I_{p\dot{q}} & \; , &\qquad
  u_{ {a}}^{I}  \tilde{\sigma}^{ {a} {\alpha} {\beta}} =- 2 v_{\dot q}^{-( {\alpha}}\gamma^I_{q\dot{q}}v_{{q}}^{+}{}^{ {\beta})}\; , \quad
\end{eqnarray}
where $\gamma^I_{p\dot{q}}=:\tilde{\gamma}^I_{\dot{q}p}$ are Klebsh-Gordan coefficients of
SO(8) group, $q,p=1,..., 8$ are  s-spinor (8s) indices, $\dot{q},\dot{p}=1,...,8$ are c-spinor (8c) indices  and I=1,.., 8 is SO(8) vector  index (8v-index). The above relations involve the spinor frame variables and also the elements of
the inverse of the spinor moving frame matrix \eqref{harmV-1=10D} the blocks of which obey
\eqref{v-qv+p=10D}.

Among the consequences of the above constraints, let us notice
$$
(v^-_{\dot{q}}\Gamma_\mu)_\alpha= u_\mu^= v_{\alpha\dot{q}}^+ - u_\mu^I v_{\alpha p}^-\gamma^I_{p\dot{q}}\; , \qquad (v^-_{{q}}\tilde{\Gamma}_\mu)^\alpha= u_\mu^= v^{+ \alpha }_{q} + u_\mu^I\gamma^I_{q\dot{p}} v^{-\alpha} _{\dot{p}}\; , \qquad
$$
which imply
$$
v^-_{\dot{q}}\Gamma_{\mu\nu}v^-_{{p}}= 2u_{[\mu}^= u_{\nu]}^I \gamma^I_{p\dot{q}}\; .
$$

\subsection{Complex spinor frame variables in D=10}

The internal  vector  frame
\begin{eqnarray}\label{UinSO8}
  U_I^{(J)}=\left(U_I{}^{\check{J}}, \frac 1 2 \left( U_I+ \bar{U}_I\right), \frac 1 {2i} \left( U_I- \bar{U}_I \right)\right) \; \in \; SO(8)
 \;  \qquad
\nonumber \\ \Rightarrow \qquad
\begin{cases} U_IU_I=0\; , \qquad \bar{U}_I\bar{U}_I=0
\; , \qquad U_I\bar{U}_I=2\; , \cr
U_IU_I{}^{\check{J}}=0\; , \qquad \bar{U}_IU_I{}^{\check{J}}=0
\; , \qquad U_I{}^{\check{J}}U_I{}^{\check{K}}=\delta ^{\check{J}\check{K}} \;  \end{cases}
\end{eqnarray}
is related to the   s-spinor and c-spinor frames \eqref{bwwSO8}
by
\begin{eqnarray}
 \label{Ug8=bwwC}
&& U\!\!\!\!/{}_{q \dot{p}}:=  \gamma^I_{q \dot{p}} U_I = 2  \bar{w}_{{q}A} w_{\dot{p}}^A  \; ,  \qquad
 \bar{U}\!\!\!\!/{}_{q \dot{p}}:= \gamma^I_{q \dot{p}} \bar{U}_I = 2   w_q^{A} \bar{w}_{\dot{p}A} \; , \qquad \\
\label{U1=bwgw}
&&  U_I\delta_A{}^B =  \bar{w}_{{q}A}\gamma^I_{q\dot{p}}  w_{\dot{p}}^B\; , \qquad  \bar{U}_I\delta^A{}_B =  {w}_{{q}}^A\gamma^I_{q\dot{p}} \bar{w}_{\dot{p}B} \;  \qquad
\end{eqnarray}
and
\begin{eqnarray}
\label{Ug86=wg6w+cc}
&& U\!\!\!\!/{}^{\check{J}}_{q \dot{p}}:= \gamma^I_{q \dot{p}} U_I^{\check{J}}= i w_q^{A}\sigma^{\check{J}}_{AB}  w_{\dot{p}}^{B} + i \bar{w}_{qA}\tilde{\sigma}{}^{\check{J}AB } \bar{w}_{\dot{p}B} \; , \qquad \\
\label{Us=bwgbw}
&& i \sigma^{\check{J}}_{AB}  U_I^{\check{J}} = \bar{w}_{{q}A}\gamma^I_{q\dot{p}} \bar{w}_{\dot{p}B}   \; ,  \qquad
i \tilde{\sigma}{}^{\check{J}AB } U_I^{\check{J}}= {w}_{{q}}^A\gamma^I_{q\dot{p}} w_{\dot{p}}^{B} \; . \qquad
\end{eqnarray}
Here $ \check{I}=1,\ldots, 6$, $A,B,C,D=1,\ldots , 4$ and
\begin{eqnarray}\label{sigma6d=}
{\sigma}^{\check{I}}_{AB}= - {\sigma}^{\check{I}}_{BA}= - (\tilde{\sigma}^{\check{I} AB})^*=  {1\over 2}\epsilon_{ABCD}
\tilde{\sigma}^{\check{I}\, CD}  \qquad
\qquad
\end{eqnarray}
are 6d Clebsch-Gordan coefficients which obey
\begin{eqnarray}\label{Cliff6d} {\sigma}^{\check{I}}\tilde{\sigma}^{\check{J}}+ {\sigma}^{\check{J}}\tilde{\sigma}^{\check{I}}= 2{\delta}^{\check{I}\check{J}}
\delta_A{}^B \; , \qquad  \label{so(6)id} {\sigma}^{\check{I}}_{AB}\tilde{\sigma}^{\check{I} CD}= -4
\delta_{[A}{}^{C}\delta_{B]}{}^{D}\; , \qquad {\sigma}^{\check{I}}_{AB}\, {\sigma}^{\check{I}}_{CD} = -2\epsilon_{ABCD}\; .
\qquad
\end{eqnarray}

One can use the internal spinor harmonics \eqref{bwwSO8} to form the complex Lorentz harmonics
\begin{eqnarray}\label{v-A:==}
 v_{\alpha A}^{-}:= v_{\alpha q}^{-} \bar{w}_{qA}\; , \qquad \bar{v}{}_{\alpha}^{-A}:= v_{\alpha {p}}^{-} {w}_{{p}}^{\; A} \; , \qquad v_{\alpha A}^{+}:= v_{\alpha \dot{p}}^{+} \bar{w}_{\dot{p}A}\; , \qquad  \bar{v}{}_{\alpha}^{+A}:= v_{\alpha \dot{p}}^{+} {w}_{\dot{p}}^{\; A} \; ,  \qquad \\
\label{v-A:==-1}   v_{ A}^{-\alpha}:= v_{\dot{q}}^{-\alpha} \bar{w}_{\dot{q}A} , \qquad \bar{v}{}^{-A\alpha }:= v_{\dot{q}}^{-\alpha} {w}_{\dot{q}}^{\; A} , \quad v_{A}^{+\alpha }:= v_{q}^{+\alpha } \bar{w}_{qA}, \qquad \bar{v}{}^{+A\alpha}:= v_{{q}}^{+\alpha} {w}_{{q}}^{\; A} . \qquad
\end{eqnarray}

Using the above properties of the internal harmonics, especially $\bar{w}_{pA}\gamma^{I}_{p\dot q}w_{\dot q}{}^B= U^I \delta_A{}^B$,
we find that the above equations imply the 10D counterpart of Eq. \eqref{v-AG2v-B=kU} (\eqref{lAG2lB=11D})):
 $$ v^{-A \alpha}\Gamma_{\mu\nu\; \alpha}{}^{\beta }v_{\beta B}^{-}= u_{[\mu}^= U_{\nu]}^I\delta^A{}_B\; . \qquad $$

\subsection{Cartan forms and derivatives of spinor frame variables/Lorentz harmonics}
\label{CartanF}

The derivatives  of the vector frame variables (vector harmonics) which respect the constraints \eqref{Uab=in10D}
are expressed in terms of $SO(1,D-1)$ Cartan forms $\Omega^{= I}:=u_a^{=}du^{aI}$,  $\Omega^{\# I}:=u_a^{\#}du^{aI}$,  $\Omega^{(0)}:={1\over 4}u_a^{=}du^{a\#}$ and $\Omega^{IJ}:=u_a^{I}du^{aJ}$ by (see \cite{Bandos:2017eof} and references therein):
\begin{eqnarray}
\label{Du--}  Du^{=}_a &:=& du^{=}_a + 2u^{=}_a \Omega^{(0)} = u^{I}_a \Omega^{=I}
\; , \qquad \\ \label{Du++}  Du^{\#}_a &:=&d u^{\#}_a - 2u^{\#}_a \Omega^{(0)} =
u^{I}_a \Omega^{\# I} \; , \qquad \\ \label{Dui}  Du^{I}_a &:=& du^{I}_a + u^{J}_a
\Omega^{JI} = {1\over 2} u^{\#}_a \Omega^{= I} + {1\over 2} u^{=}_a \Omega^{\# I} \; .
\qquad
\end{eqnarray}

As $Spin (1,D-1)$, the double covering of the Lorentz group $SO(1,D-1)$, is locally isomorphic to it,
the tangent space to  $Spin (1,D-1)$ is isomorphic to tangent space to  $SO(1,D-1)$.
Hence the  derivatives of  spinor frame variables (spinor harmonics) are also expressed in terms of the above  Cartan forms.

For D=10 one finds (see \cite{Bandos:2017eof} and refs therin)
\begin{eqnarray}
\label{Dv-dq=} &  Dv_{\alpha {q}}^{\; -} := dv_{\alpha {q}}^{\; -}+ \Omega^{(0)} v_{\alpha {q}}^{\; -} + {1\over 4} \Omega^{IJ}
v_{\alpha {p}}^{\; -}\gamma_{{p}{q}}^{IJ} = {1\over 2} \Omega^{=I}  \gamma_{q\dot{q}}^{I} v_{\alpha \dot{q}}^{\;+} \; , \qquad
\\
 \label{Dv+q=} & Dv_{\alpha \dot{q}}^{\;+}   :=  dv_{\alpha \dot{q}}^{\;+}    -
\Omega^{(0)} v_{\alpha \dot{q}}^{\;+}   + {1\over 4} \Omega^{IJ}  v_{\alpha \dot{p}}^{\;+}  \tilde{\gamma}_{\dot{p}\dot{q}}^{IJ} =  {1\over 2}
 \Omega^{\# I} v_{\alpha {q}}^{\; -}  \gamma_{q\dot{q}}^{I}\; , \qquad
\end{eqnarray}
and
\begin{eqnarray}
\label{Dv-1-q} &  Dv_{\dot{q}}^{-\alpha} := dv_{\dot{q}}^{-\alpha} + \Omega^{(0)} v_{\dot{q}}^{-\alpha} +
{1\over 4} \Omega^{IJ} \tilde{\gamma}_{\dot{q}\dot{p}}^{IJ} v_{\dot{p}}^{-\alpha} = - {1\over 2} \Omega^{=I}
 v_{{q}}^{+\alpha} \gamma_{q\dot{q}}^{I}\; , \qquad \\
\label{Dv-1+q} &  Dv_{{q}}^{+\alpha} := dv_{{q}}^{+\alpha} - \Omega^{(0)} v_{{q}}^{+\alpha} +
{1\over 4} \Omega^{IJ} v_{{p}}^{+\alpha} \gamma_{{p}{q}}^{IJ} = - {1\over 2} \Omega^{\#I }
 \gamma_{q\dot{p}}^{I}v_{\dot{p}}^{-\alpha} \; . \qquad
\end{eqnarray}

The above equations can be used also for the case of $D=11$  spinor frame variables (spinor harmonics) if we  assume that $I,J=1,...,9$, $\; p,q=1,...,16$, identify $\dot{q}$ with $q$ and replace
 the SO(8) Klebsh-Gordan coefficients $\gamma_{p\dot{q}}^{I}$  by $16\times 16$ nine dimensional  gamma matrices   $\gamma_{p{q}}^{I}=\gamma_{qp}^{I}$.

\end{document}